\documentclass[12pt, a4paper]{article}
\usepackage{preamble_article}
\newcommand{\figpath}{\string ./results_figure/}
\newcommand{\tabpath}{\string ./results_table/}

\title{Do crises increase parochial behavior? Evidence from donations during Covid\thanks{We would like to thank the following people for helpful comments on earlier drafts: Julia Chytilova, Sarah Jacobson, Abigail Payne, Tomasso Reggiani and seminar participants at the ASSA meetings, University of Bristol and WZB Berlin Social Science Center. Thanks are also due to the Charities Aid Foundation for making their data available. As per our agreement all reported results are based on aggregated data, to ensure that no individual behavior can be identified.}}

\author{
    Esteban Jaimovich\thanks{University of Turin (ESOMAS) and Collegio Carlo Alberto. Email \texttt{esteban.jaimovich@unito.it};} 
    \qquad Sarah Smith\thanks{University of Bristol, CEPR and IFS. Email: \texttt{sarah.smith@bristol.ac.uk};}
    \qquad Derrick Xu\thanks{University of Southampton, Email: \texttt{derrick.xu@soton.ac.uk}.}
}

\date{\today}

\begin{document}

\maketitle

\begin{abstract}

Do people behave more favorably towards their in-group during a crisis? Defining in/out-groups by geography, we study donations to local versus non-local charities during the Covid pandemic. The evidence points to increased parochialism: We document a relative increase in donations to local charities and show that donors in high-Covid areas increased their local giving and became less responsive to international disasters. Increased local giving was most pronounced during the first year of the pandemic. Our interpretation is that greater parochialism in donations was due to heightened concern about the immediate impact of Covid at a time when localities were salient.  

\vspace{10pt}
\end{abstract}

%%%%%%%%%%%%%%%%%%%%%%%%%%%%%%%%%%%%%%%%%%%%%%%%%%%%%%%%%%%%%%%%%%%%%%%%%%%%%%%%%%%%

\clearpage 
\begin{quote}
``There is a Christian concept that you love your family and then you love your neighbor and then you love your community and then you love your fellow citizens and then, after that, prioritize the rest of the world.'' \\
\hspace*{\fill} US Vice President, J.D. Vance
\end{quote}

%%%%%%%%%%%%%%%%%%%%%%%%%%%%%%%%%%%%%%%%%%%%%%%%%%%%%%%%%%%%%%%%%%%%%%%%%%%%%%%%%%%%
\section{Introduction} 
\label{sec:intro}
%%%%%%%%%%%%%%%%%%%%%%%%%%%%%%%%%%%%%%%%%%%%%%%%%%%%%%%%%%%%%%%%%%%%%%%%%%%%%%%%%%%%

Whether people favor in-group members (parochialism) or give equal weight to out-group members (universalism) characterizes their attitudes on many fundamental policy issues \citep{enke2020moral, romano2021national, enke2023structure, cappelen2023universalism}. Several recent policy choices made by Western governments, for example, cuts to foreign aid, the imposition of tariffs and reduced support for climate change mitigation, could be argued to indicate increasing parochialism. A possible explanation suggested by the literature is that this increase in parochial behavior is a response to external threats \citep{choi2007coevolution, bauer2016can} --- in this case, a series of recent crises, including Covid, climate change, the war in Ukraine and the increased cost of living.\footnote{From a historical perspective, these may not be the worst ever crises. But ``crisis coverage" (i.e. media mentions of crisis) reached an all-time high in 2020 \cite{geiss2025inflation, bhatia2025crisis} and the World Economic Forum used the term, \href{https://polycrisis.org/resource/were-on-the-brink-of-a-polycrisis-how-worried-should-we-be/}{poly-crisis}, to describe the state of the world in 2023.} Consistent with this argument, this paper presents new evidence from England that the Covid pandemic was associated with an increase in parochial behavior, measured by donating to local charities versus non-local charities. 

We are not the first to study whether crises increase favoritism towards in-group members. We build on several studies in less developed countries that look at the effects of war, violent conflict and natural disasters.\footnote{\cite{bentzen2021crisis} evidences an increase in religiosity in response to Covid-19 similar to the effect of natural disasters in \cite{bentzen2019} suggesting parallels between Covid and other natural disasters.} The main take-away from this literature is that there is a positive effect on co-operation, and that this is targeted towards in-group members. Reviewing the evidence, however, \cite{bauer2016can} conclude that existing studies remain speculative on whether there is increased in-group bias because too few studies define out-groups consistently, or at all. By contrast, our study captures favorable behavior towards in-group and out-group members in a consistent way using donations to local versus non-local charities \citep{enke2020moral, enke2023structure, cappelen2023universalism, enke2024universalism}. Specifically, we define increased parochial behavior as an increase in donations to local charities, relative to donations to non-local (regional, national or international) charities. Defining in-group/out-group members by geography is in line with previous studies \citep{bauer2014war, bauer2016can, calo2017natural, enke2020moral, romano2021national, enke2023structure, cappelen2023universalism}; it is also natural in the UK context where localized geographies are central to many people's identity \citep{brace1999finding, arthur2019human}.\footnote{As evidence of localized identities, England stands out, compared to other, English-speaking countries, for the number of local dialects and accents, which can vary even within a 10- or 20-mile radius \citep{trudgill1999dialects}.} 

We draw on uniquely rich data to study the effect of Covid. First, we use population, register data for all charities in England to document that charities operating locally,\footnote{Our definition of local is based around 152 English (upper tier) local authorities (with average population 170,000).} saw an increase in donation income that was more than 50 per cent greater than that of charities that operate at the non-local (i.e. regional/national/international) level. We then turn to the charity-account records of more than 60,000 donors. Mirroring the charity-level findings, we show that there was an increase in donors' share of local giving, i.e. donations to charities that operate within their local area, relative to their total giving.\footnote{Previous studies based on experimental evidence have been inconclusive on whether donations became more parochial during the Covid pandemic. \cite{grimalda2021exposure} found an effect of personal exposure on the decision to donate a monetary bonus to a charity active in Covid-relief efforts. Donors gave more to a local charity than a national/global charity, but there was no benchmark comparison, making it unclear whether parochial behavior did increase. \cite{adena2022covid} show that local severity and media coverage increased responsiveness to a donation prompt that mentioned Covid, but found no difference according to whether the recipient was national or international. A further, non-experimental study, \cite{fridman2022increased} found that local Covid threat increased generosity, but that donations went to national, social services charities.}

What might explain an increase in local giving during a crisis such as Covid? The existing literature emphasizes two channels \citep{bauer2016can}: the first is that there may be a psychological response to a crisis that causes a long-term shift in preferences in favor of the in-group, the second is that parochial behavior may be a more temporary response to increased local need, enabling communities to deal effectively with an external threat.\footnote{For example, natural disasters are associated with spikes in donations that largely originate from residents close to the affected area \citep{deryuginamarx2021,deryugina2025disparities,meonverwimp2022,schwirplies2023}} 

The evidence on the response to Covid does not support the first channel. Both charity-level and donor-level data show that the increase in local donations, relative to non-local donations, was largely reversed after the first year of the crisis, mirroring the reduction in deaths. This is not consistent with the idea, suggested by \cite{choi2007coevolution}, of a longer-term shift in underlying preferences in favor of in-group members. Instead, the timing of the response points to parochial behavior as a response to (real or perceived) need. 

In contrast to previous studies' focus on localized crises, however, the Covid pandemic was a global crisis, raising the question of why \textit{local} giving increased. A potentially novel explanation, highlighted by our conceptual framework, is that even during a time of general crisis, concern for local need relative to non-local need may intensify, particularly in areas where in-group preferences are already strong. Consistent with this, we show that the increase in local giving was greater in areas where people felt a higher level of local belonging.     

We also present evidence in support of the idea that increased local giving was a response to (real or perceived) greater \textit{local} need. Following previous studies \citep{adena2022covid, fridman2022increased}, we separate donors by their degree of local exposure to Covid (measured by the local death rate) and show that the increase in local giving was driven by donors in areas with high Covid deaths. We plausibly rule out that the increase in local giving in high-exposure areas might be a response to particular fundraising activities by local charities by showing also that donors in these areas became less sensitive to large-scale, national fundraising campaigns for international disaster relief. Our interpretation is that the increase in local giving in high-Covid areas was due to heightened concern among donors about the immediate effects of Covid in their communities. This was a time when local communities were highly salient as a result of lockdown and travel restrictions, as well as a local focus to media reporting. As further support for this interpretation, we present evidence that more-exposed donors were more likely to worry about the immediate impacts of Covid. We think it plausible that donors in high-Covid areas perceived a greater increase in local need, relative to national and international impacts. 

If parochial behavior is a response to heightened local need, is there evidence that it serves an instrumental purpose and helps communities to deal more effectively with crises? We are not aware of any evidence on this issue.\footnote{Several, related papers study the effect of social and civic capital on the spread of Covid, but do not consider how they impact on the response to Covid \citep{bartscher2021social, borgonovi2021evolution, barrios2021civic}.} We end our study with preliminary, correlational  evidence showing that areas with the biggest increase in local giving bounced back better after the Covid pandemic in terms of well-being. This supports the idea that parochial behavior, as a response to local need, may be able to mitigate some of the damaging effects of a crisis.

The plan of the paper is as follows. Section \ref{sec:model} presents a simple conceptual framework to explain our thinking on how a crisis might affect donations to local and non-local charities. Section \ref{sec:context} provides contextual information on the Covid pandemic in England. Section \ref{sec:charity} draws on charity-level data to show a stronger response in donations to local charities following the onset of the Covid pandemic. Section \ref{sec:donor} presents our analysis of the donor-level data. Section \ref{sec:discussion} concludes. 

%%%%%%%%%%%%%%%%%%%%%%%%%%%%%%%%%%%%%%%%%%%%%%%%%%%%%%%%%%%%%%%%%%%%%%%%%%%%%%%%%%%%
\section{Conceptual framework}
\label{sec:model}
%%%%%%%%%%%%%%%%%%%%%%%%%%%%%%%%%%%%%%%%%%%%%%%%%%%%%%%%%%%%%%%%%%%%%%%%%%%%%%%%%%%%

We present a simple conceptual framework to clarify our thinking on the effect
of Covid on the optimal allocation of donations across local and non-local
charities. There are $N$ groups corresponding to different geographic areas
and $N$ charities, each of which is located and operates exclusively within
one area.\footnote{To keep the model brief we consider a setup where charities
operate at the local level and whose areas of operation never overlap. The
model could be easily extended to encompass charities that operate beyond the
local level, such as national or international level.} Donor $i$ belongs to
group $I(i)\in\mathcal{I}=\left\{  1,2,...,N\right\}  $. $I(i)$ also denotes
donor $i $'s local charity. The remaining $N-1$ charities $j\neq I(i)$ are
non-local from donor $i$'s perspective.

In each period $t,$ donor $i$ chooses private consumption $c_{i,t}$ and
donations across the $N$ different charities $\{x_{i,t,j}\}_{j=1}^{N}$ to
maximize:%
\begin{equation}
U_{i,t}=c_{i,t}+\alpha_{i,I}\cdot\left(  \frac{x_{i,t,I(i)}^{\left(
1-\sigma/\eta_{t,I(i)}\right)  }}{1-\sigma/\eta_{t,I(i)}}\right)
+\alpha_{i,O}\cdot\sum\nolimits_{j\neq I(i)}\left(  \frac{x_{i,t,j}^{\left(
1-\sigma/\eta_{t,j}\right)  }}{1-\sigma/\eta_{t,j}}\right)  ,\label{Utility i}%
\end{equation}
subject to the budget constraint
\begin{equation}
c_{i,t}+\sum\nolimits_{j=1}^{N}x_{i,t,j}=y_{i,t},\label{budget constraint}%
\end{equation}

The utility function (\ref{Utility i}) includes two sets of parameters
governing warm-glow motives. First, are in-group vs.\thinspace\thinspace
out-group weights: $\alpha_{i,I}$ and $\alpha_{i,O}$, which capture donor
$i$'s intrinsic attachment to local and non-local charities, respectively. We
impose $\alpha_{i,I}\geqslant\alpha_{i,O}$ for all $i$, allowing $\alpha
_{i,I}/\alpha_{i,O}$ to vary across $i$ to reflect heterogeneities in in-group
bias.\footnote{The special case in which $\alpha_{i,I}=\alpha_{i,O}$ would
reflect an individual who attaches the same weight across all existing
charities irrespective of where they operate; this would be the case of a
\emph{pure universalist} donor.} Second, $\eta_{t,I(i)}\geq1$ and
$\eta_{t,j}\geq1$ for $j\neq I(i)$, are time-varying parameters capturing
perceived need --- or salience of need --- in the local area and other areas,
respectively. A higher value $\eta_{t,I(i)}$ (resp. $\eta_{t,j}$) reduces the
curvature of the warm-glow derived from giving to local-charity $I(i)$ (resp. to the
non-local charity $j$); i.e., donors become less quickly \textquotedblleft
satiated\textquotedblright\ in their warm-glow from giving to a charity
operating in an area with higher perceived need. We let $0<\sigma<1$,
reflecting the decreasing marginal utility of warm-glow giving and that increases in $\eta_{t,j}$ raise the marginal utility of giving to $j$.

We summarize donor $i$'s optimal allocation in terms of relative donations
across different areas/charities.\ This makes transparent how (i) in-group vs.
out-group weights and (ii) area-specific need shape the split of
giving. 

Comparing the first-order conditions for the local charity $I(i)$ and
a generic non-local charity $j\neq I(i)$ yields:%
\begin{equation}
\frac{\left(  x_{i,t,I(i)}^{\ast}\right)  ^{1/\eta_{t,I(i)}}}{\left(
x_{i,t,j}^{\ast}\right)  ^{1/\eta_{t,j}}}=\left(  \frac{\alpha_{i,I}}%
{\alpha_{i,O}}\right)  ^{\frac{1}{\sigma}}.\label{optimal allocation local}%
\end{equation}
Equation (\ref{optimal allocation local}) shows that local giving is higher
relative to that to any non-local charity when the donor's in-group weight is larger
(higher $\alpha_{i,I}/\alpha_{i,O}$). It also shows that donations respond to
the relative intensity of perceived-need: holding $\alpha_{i,I}$ and
$\alpha_{i,O}$ fixed, an increase in $\eta_{t,I(i)}$ that is greater than the
increase in $\eta_{t,j}$ raises local giving relative to $j$, and conversely
when the increase in $\eta_{t,j}$ is the larger one.

Second, comparing the first-order conditions across two non-local charities
yields:%
\begin{equation}
\frac{\left(  x_{i,t,j}^{\ast}\right)  ^{1/\eta_{t,j}}}{\left(  x_{i,t,k}%
^{\ast}\right)  ^{1/\eta_{t,k}}}=1.\label{optimal allocation non-local}%
\end{equation}
Equation (\ref{optimal allocation non-local}) implies that the optimal split
across non-local charities depends only on their relative perceived-need
across $j$ and $k$: if $\eta_{t,j}=\eta_{t,k}$ then donations are equalized
across $j$ and $k$; if $\eta_{t,j}>\eta_{t,k}$ the donor tilts donations
toward $j$ at the expense of $k$.

Comparing (\ref{optimal allocation local}) with
(\ref{optimal allocation non-local}) highlights a key asymmetry in how
donations respond to increases in perceived need. Consider a general calamity
(such as the Covid\ pandemic) that raises perceived need in \emph{all} areas,
captured by a common increase in all $\left\{  \eta_{t,j}\right\}  _{j=1}^{N}$. Equation (\ref{optimal allocation non-local}) implies that such a uniform
shock leaves the relative allocation across non-local charities unchanged. By
contrast, equation (\ref{optimal allocation local}) implies that
the\emph{\ local vs. non-local} ratio shifts toward the local charity whenever
$\alpha_{i,I}>\alpha_{i,O}.$

To see this, suppose that before the crisis $\eta_{t-1,I(i)}=\eta_{t-1,j}=1$,
and the crisis leads to $\eta_{t,I(i)}=\eta_{t,j}=1+\kappa$, with $\kappa>0.$
Then, equation (\ref{optimal allocation local}) implies that the
\emph{local/non-local} \emph{donation ratio }rises from $(x_{i,t-1,I(i)}%
^{\ast}/x_{i,t-1,j}^{\ast})=\left(  \alpha_{i,I}/\alpha_{i,O}\right)
^{1/\sigma}$ to $(x_{i,t,I(i)}^{\ast}/x_{i,t,j}^{\ast})=\left(  \alpha
_{i,I}/\alpha_{i,O}\right)  ^{\left(  1+\kappa\right)  /\sigma}$. That is, in
periods of greater social upheaval, a common rise in perceived need amplifies
the impact of \emph{pre-existing} in-group bias. Moreover, this increase is
stronger for donors with greater in-group bias (i.e., those with a larger
$\alpha_{i,I}/\alpha_{i,O}$).

The framework therefore illustrates how the interaction between perceived need
and in-group preferences can generate an increase in the share of donations
directed to local charities. Donors respond not only to relative increases in
local need, but also --- when $\alpha_{i,I}>\alpha_{i,O}$ --- to aggregate
increases in perceived need that amplify the attractiveness of local giving
more than non-local giving. We explore these implications in the empirical
analysis below.

\section{Covid in England}
\label{sec:context}
%%%%%%%%%%%%%%%%%%%%%%%%%%%%%%%%%%%%%%%%%%%%%%%%%%%%%%%%%%%%%%%%%%%%%%%%%%%%%%%%%%%%

England experienced among the highest Covid death rates in Europe, just below that in the US.\footnote{https://data.who.int/dashboards/covid19/deaths} Most of the deaths came in two peaks in Spring 2020 and early 2021 and these coincided with two, national lockdown periods (a third, national lockdown took place in November 2020).\footnote{Figure \ref{fig:timeline} in the Appendix shows the timeline of deaths and lockdowns during the pandemic. One of the advantages of studying Covid in England is that almost all lockdowns (outside a short period of local lockdowns in July 2020) were national allowing us to separate the effect of local, high exposure from the effect of lockdowns. Figures \ref{fig:national_lockdown_on_local_nat} -- \ref{fig:change_in_local_giving_following_local_lockdowns} in the Appendix are consistent with the idea that donations responded to Covid deaths, rather than lockdowns.} We follow \cite{adena2022covid} and study the effect of local exposure, measured by local authority death rates. We categorize high/low exposure areas based on above/below-median cumulative death rates to December 2022 --- these areas are shown in Appendix Figure \ref{fig:ulad_high_low_covid}. 

The Covid pandemic created an economic crisis as well as a health crisis. GDP fell by 19.4 per cent in the second quarter of 2020, largely as a result of lockdowns and other restrictions. The magnitude of local economic impact, measured by above/below-median loss in GDP (see Appendix Figure \ref{fig:ulad_high_low_gdp}) was uncorrelated with local health exposure (high/low death rates), allowing us to separate health and economic impacts. We show below that increased local giving is correlated with local health impacts, not local economic impacts. Analysis of UK news content reveals that health stories dominated economic stories during the pandemic, suggesting that the health impacts were likely to be most salient at the time (Appendix Figure \ref{fig:time_series_news_share_by_topic}).

A priori, the idea that Covid would increase parochial behavior is not obvious: It was a global crisis not a local one. However, several factors are likely to have intensified the localized experience of the pandemic. Local communities became more important in everyday lives as travel opportunities reduced. Covid increased the benefits from acts of parochial altruism --- for example, wearing masks protected people you came into contact with locally, while many people helped neighbors when they had to self-isolate. Local identities are traditionally strong in many parts of England and, at a time of uncertainty, people may have valued the known and familiar.\footnote{\cite{stewart2023group} show how shocks may be associated with increased stereotyping and there was documented increased hostility to certain out-group members, for example, an increase in hate crime against members of the East-Asian communities \citep{tessler2020anxiety}} There was strong, early support for the  National Health Service (NHS), but low perceived competence of the national government may have undermined any sense of a collective, national identity. In addition, the majority of news stories during the pandemic had a local, not national/international, angle, making local impacts and needs more salient (Appendix Figure \ref{fig:time_series_local_share}). 

\begin{figure}[!htbp]
    \centering
    \includegraphics[width=0.9\textwidth]{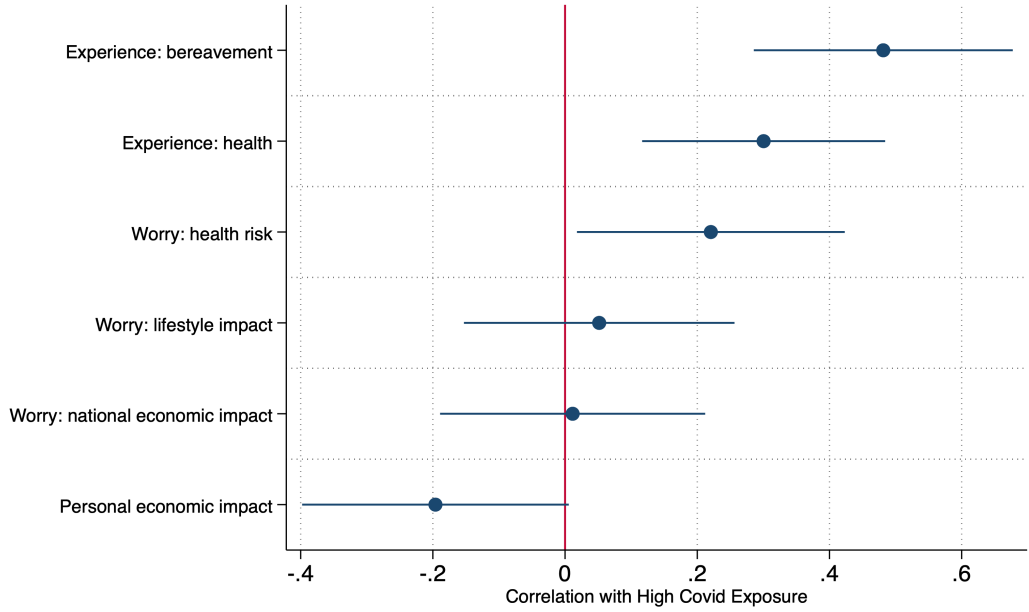}
    \caption{Correlation between Covid exposure and its potential consequences}
    \label{fig:treatment_correlation}

    \vspace{5pt} 
    \parbox{\linewidth}{\footnotesize \textit{Notes}: The figure plots correlations ($\beta$) between different, standardized outcomes (taken from the British Election Survey) and high Covid exposure from the following constituency-level regression: $ y_i = \alpha + \beta HighCovid_i + \delta'X_i + \varepsilon_i$, where $i$ denotes one of 543 constituencies in England. High Covid Exposure (0/1) indicates above-median Covid death rates. Labels of outcome variables and corresponding questions: \textit{Personal bereavement} refers to whether someone the respondent personally knew died as a result of coronavirus; \textit{Personal health impact} refers to how badly the respondent was affected by coronavirus; \textit{Worry: lifestyle impact} refers to how worried the respondent is about the impact of coronavirus on their way of life; \textit{Worry: National economic impact} refers to how worried the respondent is about the impact of coronavirus on the economy; \textit{Worry: health risk} refers to how worried the respondent is about themselves, or someone close to them, catching coronavirus; \textit{Personal economic impact} refers to whether the respondent’s household monthly income decreased since the coronavirus outbreak. Controls include socio-economic characteristics at the constituency level, including age, female share, gross personal income, education level, UK citizenship, British identity, English identity, religiosity, share of the White population, Conservative/Labour vote share, rural share. }
\end{figure}

What might explain why local giving increased specifically in areas of high Covid exposure? Our interpretation is \textit{not} that people responded to higher deaths per se, but that people in high-exposure areas experienced heightened concern about the impacts of Covid in their communities. Figure \ref{fig:treatment_correlation} lends support to this argument. It shows correlations between living in a high-exposure areas and different measures of personal experience of, and worry about, the impacts of Covid: Living in an area with high-Covid exposure was associated with increased experience of, and worry about, the immediate impacts of Covid (knowing someone who died, being affected by Covid, worrying about the health risk) not with general concern about the national, macro consequences.\footnote{This idea is consistent with the finding of \cite{campante2024virus} that anxiety in the US associated with Ebola cases led to increased anti-immigration sentiment.}   

%%%%%%%%%%%%%%%%%%%%%%%%%%%%%%%%%%%%%%%%%%%%%%%%%%%%%%%%%%%%%%%%%%%%%%%%%%%%%%%%%%%%
\section{Charity-level analysis}
\label{sec:charity}
%%%%%%%%%%%%%%%%%%%%%%%%%%%%%%%%%%%%%%%%%%%%%%%%%%%%%%%%%%%%%%%%%%%%%%%%%%%%%%%%%%%%

We begin our empirical analysis by examining the impact of Covid on the evolution of donations to local and non-local charities using charity-level data.

%===================================================================================
\subsection{Charity-level data}
%===================================================================================

All charities registered in England and Wales are legally required to file an annual return of their income and spending to the Charity Commission. The level of detail they have to report depends on the charity's income: All charities report total income and spending; charities with a gross income exceeding £500,000 report sub-categories of income (including donations) and of spending (including fundraising expenditure). Our analysis therefore focuses on these larger charities, comprising 13\% of registered charities but approximately 91\% of total sector income during the study period.\footnote{A peculiarity of the charitable sector in England and Wales is that charities can freely select the start and end dates for their fiscal years. We exclude observations which are not for a full, 12-month period (approximately 5.5\% for charities with incomes above £500,000).}

The Charity Commission's registry includes information about the geographic scope of a charity's operations. We focus on charities with headquarters in England. We classify a charity as \textit{local} if it operates exclusively at the local authority level. The majority of these operate within a single local authority; 90\% operate in five or fewer local authorities. In contrast, charities are considered \textit{non-local} if their activities extend beyond this, including regional, national, or international operations. 

Local charities are, on average, smaller than non-local ones (Appendix Table \ref{tab:summary_stats_charity}). However, Appendix Table \ref{tab:charity_presence_sectors} shows that there is a similar distribution of local and non-local charities across broad sectors of activity (health, social services, etc), alleviating concerns that our results are confounded by heterogeneous sectoral responses. In the analysis below, we also confirm that the results are not attributable to differential selection of local/non-local charities into the estimation sample during the pandemic.  

%===================================================================================
\subsection{Regression results}
%===================================================================================

Table \ref{tab:charity_level_local_vs_nonlocal}, Column (1) reports results from estimating the following equation:
\begin{equation}
    \ln D_{jt(j)} = \gamma_1 \text{Post}_{t(j)} + \gamma_2 \text{Post}_{t(j)} \times \text{Local}_j + \delta \text{Year}_{t(j)} + \varsigma_j + \varepsilon_{jt(j)}.
    \label{eq:charity_reg}
\end{equation}
The dependent variable is the logarithm of total donations received by charity $j$ during the year $t(j)$. Note that time-periods $t(j)$ are charity-specific since charities can choose the month for the beginning of their fiscal year. $\text{Post}_{t(j)}$ is defined as the share of months during year $t(j)$ that occur after March 2020.\footnote{For example, if charity $k$ ran a fiscal year from April 2019 until March 2020, we would then have $\text{Post}_{20(k)}$ equal to 1/12. If charity $l$ ran a fiscal year from August 2019 until September 2020, we would instead have $\text{Post}_{20(l)}$ equal to 7/12.} In that regard, the variable $\text{Post}_{t(j)}$ captures an intensity of treatment. $\text{Local}_j$ is a dummy variable that equals one if charity $j$ is classified as local, i.e.\ it operates exclusively at the local authority level. The regression includes a time linear trend ($\text{Year}_{t(j)}$) and charity fixed effects ($\varsigma_j$). Using this specification to identify a differential effect of Covid on donations to local charities relies on a standard common trends assumption. Appendix Figure \ref{fig:event_cc} shows the results from running an event-study specification, confirming null effects prior to 2020.   

\begin{table}[!htbp]
    \centering
    \caption{Charity-level regressions: local vs non-local}
    \label{tab:charity_level_local_vs_nonlocal}
    \resizebox{0.90\textwidth}{!}{\estauto {\tabpath charity_commission_evidence.tex}{15}{c}} 

    \vspace{5pt} 
    \parbox{\linewidth}{\footnotesize \textit{Notes}: The dependent variable is the logarithm of total donations received by charity $j$ (headquartered in the upper-tier local authority $l$) during year $t$. `Postcovid' is defined as the share of months of the charity's fiscal year that occur after March 2020. Year 202(X) is defined to match the UK fiscal year, running from March 202(X) until February 202(X+1). `Local' is a dummy variable equal to 1 if charity $i$ operates exclusively at the upper-local authority level, and 0 if it operates at a broader level. Robust standard errors clustered at the upper-local authority level in parentheses. $^{*}p<0.1$ $^{**}p<0.05$; $^{***}p<0.01$}
\end{table}

Column (1) shows that donations increased above their linear trend during Covid for both local and non-local charities. However, the increase for locally-operating charities (24.5\%) was more than 50 per cent greater than that for non-local charities (15.9\%).

Column (2) adds (log) total fundraising expenditure as an additional regressor. This controls for the possibility that local charities may have responded differently to the Covid pandemic with increased fundraising effort. As expected, fundraising expenditure has a positive coefficient, but the estimate of $\gamma_2$ is essentially unaffected.

Column (3) looks separately at years 2020 and 2021 (defined from March to Feb of the following year). Both the increase in donations for non-local charities and the additional increase in donations for local charities are stronger in the first year of the pandemic than in the second year. In the second year of the pandemic (March 2021 -- Feb 2022), the magnitude of the \textit{Local} interaction term reduces considerably and is no longer significant in all specifications. Event study analysis (Appendix Figure \ref{fig:event_cc}) confirms the absence of a significant differential effect prior to 2020.  

Columns (4 -- 6) add further controls. Column (4) includes (log) income per capita in the local authority where the charity's headquarters are located. In (5) we allow for heterogeneous linear time trends between local and non-local charities. Lastly, in (6) we exclude charities whose headquarters are in Central London. The estimates are very similar; in (5) and (6) the increase in donations for non-local charities in 2021 is not statistically significant.

One potential concern is that the sample only comprises charities with total income above £500,000. This may introduce bias arising from differential entry into and exit from the sample by local and non-local charities. In particular, since total income and income from donations are mechanically positively correlated, charities that drop below the threshold are more likely to have experienced a decline in donations, and vice versa. To address this, Appendix Table \ref{tab:charity_level_imputed} imputes donation income for charities in periods when their total income dips below the £500,000 threshold, using their average ratio of (observed) donations to total income. This increases the number of observations in column (1) from 68,873 to 94,973. Reassuringly, the estimated coefficients are close to those in Table \ref{tab:charity_level_local_vs_nonlocal}. 

%%%%%%%%%%%%%%%%%%%%%%%%%%%%%%%%%%%%%%%%%%%%%%%%%%%%%%%%%%%%%%%%%%%%%%%%%%%%%%%%%%%%
\section{Donor-level analysis}
\label{sec:donor}
%%%%%%%%%%%%%%%%%%%%%%%%%%%%%%%%%%%%%%%%%%%%%%%%%%%%%%%%%%%%%%%%%%%%%%%%%%%%%%%%%%%%

%===================================================================================
\subsection{Donor-level data}
%===================================================================================

We now turn to donor-level data, drawn from anonymized donor accounts administered by the Charities Aid Foundation (2014), henceforth CAF.\footnote{The accounts are dedicated checking accounts for making donations. Anyone can set up an account with a minimum £100 one-off payment or £10 monthly direct debit; they can make additional contributions at any time, but cannot withdraw funds. Donations can be made to any registered charity and can be made online, by phone or check. More detailed discussion of CAF accounts and the evidence to be discussed in this section is available in Appendix \ref{asec:data}.} We analyze donations made via the accounts from March 2015 -- February 2022. Over the period, 64,417 account-holders made at least one donation and in total, we observe more than 5 million donations to more than 44,000 charities. For each donation, we observe the date, the amount donated and the charity. We aggregate to the donor-year level, where year is defined from March -- February. The mean (median) number of donations per donor/year is 17 (8), while the mean (median) value of total donations is £2,205 (£590).\footnote{Donations through CAF accounts represent six per cent of total giving in the UK. The benefits of setting up a CAF account --- making it easier to obtain the tax benefits of donating under the UK Gift Aid system and to help people manage their giving --- are more important to people who give more. Appendix Table \ref{tab:ukhls_caf} and Table \ref{tab:donation_deciles} compare donations by CAF account holders with donations made by donors in a random sample drawn from the UK population.} 

We use donors' postcodes to match additional variables, including a measure of household wealth (at the six digit postcode level) and local area characteristics (urbanization, demographics, socio-economic characteristics). Appendix \ref{asec:vardata} provides full details. We also use donor postcodes to refine the geographic dimension of giving. We define \textit{local giving} as donations made to local charities (defined as above) whose area of operation lies within 25 km of the donor's address. These donations represent 75\% of the total value of donations to local charities (see Appendix Table \ref{tab:donation_shares_by_distance}).\footnote{Appendix Table \ref{tab:robustness_distance_thresholds} shows that the result is robust to alternative distance (e.g., 10 and 50 km).} As reported in Appendix Table \ref{tab:summary_statistics_caf_donation}, \textit{local giving}  accounts for 17 per cent of donor/year giving according to this definition. Appendix Table \ref{tab:covariants_of_local_giving} shows that \textit{local giving} is positively correlated with being male, wealth, living in an urban area and in areas where there is a stronger sense of national identity and ethnic homogeneity. It is negatively correlated with religiosity and with age. 

%===================================================================================
\subsection{Regression results}
%===================================================================================

Table \ref{tab:main_diff_change_after_covid}, Column (1) reports results from estimating the following equation on the CAF donor data:
\begin{equation}
    y_{it} = \alpha + \beta \text{Post}_t + \phi_i + \delta \text{Year}_t + \gamma' X_{it} + u_{it}
\end{equation}
$y_{it}$ is the share of \textit{local giving} out of total donations by donor $i$ in year $t$, defined to run from start of March - end February. Results using alternative outcome variables, such as a binary indicator for whether a donor gives to a local charity, produce similar results (see Appendix Table \ref{tab:covid_on_local_giving}). $\text{Post}_t$ is a binary indicator capturing years after the onset of the Covid pandemic i.e.\ March 2020 onward. We include donor fixed effects $\phi_i$ and a linear year trend $\delta$. $X_{it}$ controls for donor- and year-varying factors. These controls include donors' total annual giving, which captures general changes in income and spending, as well as indicators for the charitable sectors to which they give (health, social services, religious, international, environment, and other). The sector indicators account for the fact that some sectors may be more localized than others and that Covid may have impacted sectoral allocations.\footnote{Appendix Table \ref{tab:change_in_sectoral_giving} shows results for post-Covid changes in donations by sector.}

The results, reported in Column (1) mirror the charity-level analysis, confirming that Covid was associated with an increase in the share of donations made to local charities, by donors living locally. Appendix Figure \ref{fig:event_study_year_only} presents the results from event study analysis confirming the absence of a trend in local giving prior to 2020. Further analysis (Appendix Table \ref{tab:change_in_local_giving_by_cause}) shows that \textit{local giving} increased across several charitable sectors, not only health,\footnote{The National Health Service, which was very salient during the early stages of the pandemic, has a network of over 230 local and specialist NHS charities.} but also social services, environment and other.  

\begin{table}[t]
    \centering
    \caption{Donor-level regressions: change in local giving after Covid}
    \label{tab:main_diff_change_after_covid}
    \resizebox{\textwidth}{!}{\estauto {\tabpath main_diff_change_after_covid.tex}{15}{c}} 

    \vspace{5pt} 
    \parbox{\linewidth}{\footnotesize \textit{Notes}: The outcome variable is the share of a donor’s annual donations going to charities within 25 km of their location. A year runs from March to February to align with the outbreak of the pandemic. The data cover March 2015 to February 2022. \textit{Post Covid} indicates years since March 2020. \textit{High Covid} indicates upper-tier local authorities (ULADs) with cumulative Covid death rates up to 2022 at or above the median. \textit{Large GDP Shock} indicates ULADs with the change in mean annual GDP after Covid at or below the median. \textit{Strong Local Belonging} indicates constituencies where the average sense of belonging to the local community is at or above the median. All regressions include donor fixed effects, year trends, sectoral composition, total donations. Columns (2--6) include interactions of \textit{Post Covid} with gender, wealth, urbanization, the shares of people with higher education, claiming UK identity, with a religious affiliation, aged 65 or above, identifying as British White, and Conservative vote share. Column (5) includes region by year fixed effects; \textit{Year = 2020} and \textit{Year = 2021} in Column (6) capture the first and second Covid years. Standard errors are clustered at the ULAD level. $^{*}p<0.1$ $^{**}p<0.05$; $^{***}p<0.01$.}
\end{table}

Following previous studies \citep{adena2022covid, fridman2022increased}, we test for differences in response by degree of Covid exposure: 
\begin{equation}
    y_{it} = \alpha + \beta_1 \text{Post}_t + \beta_2 \text{Post}_t \times \text{High}_i + \phi_i + \delta \text{Year}_t + \gamma' X_{it} + u_{it}
\end{equation}
where $\text{High}_i$ indicates if the donor lives in a local authority with above-median death rate. The results in Column (2) show that the increase in local giving is driven by donors in high-exposure areas.\footnote{Results with a continuous death rate are shown in the Appendix Table \ref{tab:main_diff_change_after_covid_contT}.} Of course, the impact of Covid was not random and there might be other factors, associated with above-average Covid death rates that drove the increase in local giving. However, we show that the main result is robust to including a rich set of controls for the possibility of different trends in high-exposure areas: We interact the post-Covid indicator with the local population share of 65-plus, degree-educated, British White, UK identity, Conservative vote share, urbanization, religious affiliation and donor gender. Event study analysis (Appendix Figure \ref{fig:event_study_high_covid}) confirms that a differential ``high Covid'' effect emerges only in 2020. The effect is also robust to the inclusion of region-time fixed effects (Column (5)).  

Column (3) adds an interaction with a measure of local economic impact, i.e. exposure to an above-median decline in local GDP. The effect of this is negative, i.e. the share of \textit{local giving} reduces, but the interaction with high-Covid exposure is unaffected. In line with the model in Section \ref{sec:model}, Column (4) adds an interaction term with an area-level measure of strong local belonging. This has the expected positive correlation, i.e. the increase in \textit{local giving} is greater in areas where donors display greater in-group bias. We obtain similar results in Appendix Table \ref{tab:community_on_local_giving}, using other measures of local cohesion (low hate crime, high ethnic homogeneity). Finally, Column (6) shows separate effects by year: 2020 (Mar 2020 - Feb 2021) and 2021 (Mar 2021 - Feb 2022). In line with the charity-level results, there is a greater increase of \textit{local giving} in high-Covid areas during the first year of the pandemic. In 2021, the interaction term with \textit{High Covid} is positive but statistically insignificant. 

%===================================================================================
\subsection{Reduced sensitivity to international disaster appeals}
%===================================================================================

In this section we present further evidence consistent with more parochial behavior; namely that donors in high-Covid areas become less responsive to general fundraising campaigns launched nationally in the aftermath of international disasters.\footnote{This finding is at odds with \cite{meon2022pro} who find that donations to an international fundraising campaign increase in response to a local natural disaster. A key difference in their setting is that there is a fundraising campaign for victims of the local disaster, which may increase the salience of other causes, consistent with the findings of \cite{scharf2022lift}} This finding helps to address a potential concern that the greater increase in local giving in high-Covid areas may be driven by specific fundraising activities by local charities in these areas. It also rules out that increased donations to local charities represents a shift to giving to charities that are seen as more agile and better-placed to respond to need, compared to larger, national charities. The DEC appeals are from charities that are well-placed to respond to a clear and immediate need.

Following \cite{scharf2022lift}, we estimate donor responses to appeals by the UK Disasters Emergency Committee (DEC). This is an umbrella organization of fifteen large international charities\footnote{They are Action Against Hunger, ActionAid, Age International, British Red Cross, CAFOD, CARE International, Christian Aid, Concern Worldwide, International Rescue Committee, Islamic Relief, Oxfam, Plan International, Save the Children, Tearfund, and World Vision.} that co-ordinates relief in response to major, overseas natural and humanitarian disasters. In the immediate aftermath of a large-scale disaster, DEC makes a prominent, national fundraising appeal for donations. Over the period 2015--2022, there were 10 DEC appeals, including four in the post-Covid period (details in Table \ref{tab:differential_response_to_decappeal}). Note that the last two fall after the period analyzed above.

We estimate the following regression on donor-month data:
\begin{equation}
\begin{split}
    y_{it} &= \alpha + \beta_1 \text{Post}_t + \beta_2 \text{Post}_t \times \text{High}_i \\ 
    &\quad + \gamma_1 \text{DEC}_t + \gamma_2 \text{DEC}_t \times \text{Post}_t + \gamma_3 \text{DEC}_t \times \text{High}_i + \gamma_4 \text{DEC}_t \times \text{Post}_t \times \text{High}_i \\
    &\quad + \phi_i + \delta \text{Year}_t + \eta_{m(t)} + \sigma' X_{it} + u_{it}
\end{split}
\label{eq:dec_regression}
\end{equation}
where $y_{it}$ is a binary indicator equal to one if the donor gives to DEC or any of its fifteen member charities and $\text{DEC}_t$ is an indicator for the ``post-appeal'' window (defined as the first three months after the launch of an appeal). $\gamma_1$ measures the increase in international aid donations in this window. $\gamma_2$ captures the differential response to DEC appeals following Covid. The positive sign indicates that donors were more likely to respond to a DEC appeal after the pandemic; we do not interpret this as an effect of Covid since the nature of the pre/post appeals varies considerably. In particular, the response to the Ukraine appeal was the largest in DEC's forty-year history. $\gamma_4$ is the main coefficient of interest, capturing whether the change in post-pandemic response to DEC appeals varied between high/low Covid areas. We control for donor fixed effects $\phi_i$, year trend $\delta$, month fixed effects $\eta_m$, and total donations $X_{it}$ in a given month by a donor. The negative coefficient (in Panel B of Table \ref{tab:differential_response_to_decappeal}) indicates that donors in high-Covid areas became less likely to respond to DEC appeals, post-pandemic, compared to donors in low-Covid areas. This negative differential was common across all four, post-Covid appeals. 

\begin{table}[!htbp]
    \centering
    \caption{Post-pandemic responses to international appeals}
    \label{tab:differential_response_to_decappeal}

    % --- Panel A  ---
    \noindent\makebox[\linewidth][c]{\textbf{Panel A:} Post-pandemic DEC appeals}
    
    \vspace{3pt}
    {\footnotesize
    \begin{tabular*}{0.97\linewidth}{@{\extracolsep{\fill}}p{0.55\linewidth}ll}
        \toprule
        Appeal Event & Launch Date & Est. Amount Raised \\
        \midrule
        Covid Appeal was launched to help people in some of the world's most fragile places: DR Congo, Somalia, South Sudan, Yemen, Syria, Afghanistan, and Bangladesh. & 14 Jul 2020 & £62 Million \\
        \addlinespace
        Afghanistan Crisis Appeal & 15 Dec 2021 & £52 Million \\
        Ukraine Humanitarian Appeal            & 03 Mar 2022         & £445 Million \\
        Pakistan Floods Appeal    & 01 Sep 2022       & £50 Million \\
        \bottomrule
    \end{tabular*}}

    % --- Panel B ---
    \vspace{5pt} 
    \noindent\makebox[\linewidth][c]{\textbf{Panel B:} Responses to DEC appeal}
    
    \vspace{3pt}
    \resizebox{\linewidth}{!}{\estauto{\tabpath differential_response_decappeal.tex}{15}{c}}

    \vspace{5pt} 
    \parbox{\linewidth}{\footnotesize \textit{Notes}: Panel A provides an overview of the four post-pandemic DEC Appeals, collected from the DEC website on 6 February 2026. Panel B reports the coefficients. The outcome is a binary indicator equal to 1 if the donor gives to any DEC Appeal Partner charities in month $t$ (0 otherwise). $\text{DEC Appeal}$ equals one for the three months following each appeal launch. $\text{Post Covid}=1$ for months $\geq$ March 2020; $\text{High Covid}$ denotes upper tier local authorities (ULAD) classified as high Covid; All specifications include donor fixed effects, calendar-month fixed effects, a linear year trend and the donor's total donation amount in the month. Standard errors are clustered by ULAD.  $^{*}p<0.1$, $^{**}p<0.05$, $^{***}p<0.01$.}
\end{table}

%%%%%%%%%%%%%%%%%%%%%%%%%%%%%%%%%%%%%%%%%%%%%%%%%%%%%%%%%%%%%%%%%%%%%%%%%%%%%%%%%%%%
\section{Discussion}
\label{sec:discussion}
%%%%%%%%%%%%%%%%%%%%%%%%%%%%%%%%%%%%%%%%%%%%%%%%%%%%%%%%%%%%%%%%%%%%%%%%%%%%%%%%%%%%

The previous sections have provided evidence of increased parochial behavior during the Covid pandemic. Donations to local charities increased at a significantly higher rate than donations to non-local charities; analysis of CAF data confirms an increase in donations to local charities made by donors living locally. 

The framework in section \ref{sec:model} indicated several channels through which Covid might lead to an increase in donations to local charities relative to non-local charities --- increased sensitivity to local needs during a time of crisis, an increase in (perceived) local needs and a shift in preferences that puts increased weight on local needs. We discuss each in turn.    

If donors have (pre-existing) preferences that favor in-group versus out-group needs ($\alpha _{i,I}>\alpha _{i,O}$), the marginal utility of warm-glow giving to a local charity rises more sharply in response to a common need than that of any other non-local charity; moreover the strength of the differential response is increasing in the degree of in-group preference. Consistent with this, the increase in local giving was greater in areas with a strong sense of local belonging. The take-away is that (part of) the increased local giving may reflect greater sensitivity to local need during a time of general crisis.  

The finding that the increase in local giving was linked to the degree of Covid exposure is also consistent with an effect of exposure on the time-varying terms, $\eta _{t,I(i)}$ and $\eta_{t,j}$. These parameters capture (salience of) need specific to each charity's area. As previously discussed, our interpretation is that donors in high-Covid areas experienced heightened concern about the immediate impacts of Covid, at a time when local impacts were likely to be highly salient, i.e. a (temporary) relative increase in  $\eta _{t,I(i)}$ in high-Covid areas.     

The evidence does not support instead the idea that the main channel is a longer-term change in preference parameters and an increase in the weight on in-group needs ($\alpha_{i,I}$) relative to out-group needs ($\alpha _{i,O}$). Both charity-level and donor-level data show a greater increase in local giving in 2020 (Mar 2020 - Feb 2021), compared to 2021 (Mar 2021 - Feb 2022), mirroring the timing of deaths and suggesting that the increase in parochial behavior is likely to be primarily a response to perceived local need. However, we cannot rule out (small) longer-term effects altogether: Both charity-level and donor-level data point to positive local effects in 2021, while the reduced sensitivity to international disaster appeals persists at least to the end of 2022.  

The idea that parochial behavior is a response to increased need during times of crisis comes with a suggestion that it might help communities deal with the fallout from a crisis and might mitigate some of the negative effects. We are not aware of any evidence on whether this is the case; we therefore end with some preliminary, correlational evidence that parochial behavior might have helped communities to bounce back from Covid. Table \ref{tab:local_on_wellbeing} shows post-Covid changes in different measures of well-being. The coefficient on Post (2021-2022) captures the change in well-being relative to 2020, showing a significant improvement in ``yesterday happiness'' and, to a lesser extent, ``life satisfaction''. There is some evidence that the bounce back is lower in high-Covid areas, albeit insignificant. However, in high-Covid areas that saw the biggest increases in local giving, the bounce back was stronger. This finding is robust to controlling for local solidarity. While this evidence is only suggestive, it points to an important direction for future research to understand better how parochialism can cushion local communities during periods of crisis.     

\begin{table}[!htbp]
    \centering
        \caption{Effect of local giving on local well-being improvement}
        \label{tab:local_on_wellbeing}
        \resizebox{\textwidth}{!}{\estauto {\tabpath local_on_wellbeing.tex}{15}{c}}

        \vspace{5pt} 
        \parbox{\linewidth}{\footnotesize \textit{Notes}: The well-being measures are from British Election Studies, and they are asked only in the year 2020, 2021 and 2022. We make within-individual comparisons of well-being in 2021--2022 relative to 2020 (baseline). \textit{High Local Giving Rise} equals 1 if the post-Covid change in the local share of donations in a specific constituency is above the median across constituencies. Outcomes are standardized. Models include individual and post-period fixed effects. In even columns, we additionally control for differential post-2021 changes by \textit{Strong Local Belonging} (median-or-above). Standard errors are clustered at the level of constituency.}
\end{table}

We motivated this study with reference to increased parochialism in policy-making. The evidence in this paper is not conclusive that a wider policy shift can be attributed to a sense of crisis. Nevertheless, the evidence that parochial behavior increased during the Covid pandemic is consistent with this argument. We have also shed some light on what might underlie more parochial behavior in response to crises: The evidence strongly suggests that it is not driven by a longer-term shift in preferences in favor of in-group members, but a response to a greater sense of, and awareness of, local need. Exploring responses to other crises could be a useful avenue for further research. 

%%%%%%%%%%%%%%%%%%%%%%%%%%%%%%%%%%%%%%%%%%%%%%%%%%%%%%%%%%%%%%%%%%%%%%%%%%%%%%%%%%%%
\clearpage
\bibliography{references.bib}  

@techreport{deryugina2025disparities,
  title       = {Disparities in Aid for Natural Disasters},
  author      = {Deryugina, Tatyana and Marx, Benjamin M.},
  institution = {University of Illinois and Georgia State University},
  type        = {Preliminary Draft},
  month       = {March},
  year        = {2025}
}

@article{meon2022pro,
  title={Pro-social behavior after a disaster: Evidence from a storm hitting an open-air festival},
  author={M{\'e}on, Pierre-Guillaume and Verwimp, Philip},
  journal={Journal of Economic Behavior \& Organization},
  volume={198},
  pages={493--510},
  year={2022},
  publisher={Elsevier}
}

@article{barrios2021civic,
  title={Civic capital and social distancing during the Covid-19 pandemic},
  author={Barrios, John M and Benmelech, Efraim and Hochberg, Yael V and Sapienza, Paola and Zingales, Luigi},
  journal={Journal of public economics},
  volume={193},
  pages={104310},
  year={2021},
  publisher={Elsevier}
}

@article{bartscher2021social,
  title={Social capital and the spread of Covid-19: Insights from European countries},
  author={Bartscher, Alina Kristin and Seitz, Sebastian and Siegloch, Sebastian and Slotwinski, Michaela and Wehrh{\"o}fer, Nils},
  journal={Journal of health economics},
  volume={80},
  pages={102531},
  year={2021},
  publisher={Elsevier}
}

@article{borgonovi2021evolution,
  title={The evolution of the association between community level social capital and COVID-19 deaths and hospitalizations in the United States},
  author={Borgonovi, Francesca and Andrieu, Elodie and Subramanian, SV},
  journal={Social science \& medicine},
  volume={278},
  pages={113948},
  year={2021},
  publisher={Elsevier}
}

@article{arthur2019human,
  title={The human geography of Twitter: Quantifying regional identity and inter-region communication in England and Wales},
  author={Arthur, Rudy and Williams, Hywel TP},
  journal={PloS one},
  volume={14},
  number={4},
  pages={e0214466},
  year={2019},
  publisher={Public Library of Science San Francisco, CA USA}
}

@article{brace1999finding,
  title={Finding England everywhere: regional identity and the construction of national identity, 1890-1940},
  author={Brace, Catherine},
  journal={Ecumene},
  volume={6},
  number={1},
  pages={90--109},
  year={1999},
  publisher={Sage Publications Sage CA: Thousand Oaks, CA}
}

@book{trudgill1999dialects,
  title={The Dialects of England},
  author={Trudgill, Peter},
  year={1999},
  edition={2nd},
  publisher={Blackwell Publishers},
  address={Oxford}
}

@article{adena2022covid,
  title     = {COVID-19 and pro-sociality: How do donors respond to local pandemic severity, increased salience, and media coverage?},
  author    = {Adena, Maja and Harke, Julian},
  journal   = {Experimental Economics},
  volume    = {25},
  number    = {3},
  pages     = {824--844},
  year      = {2022},
  publisher = {Cambridge University Press \& Assessment}
}

@article{bauer2014war,
  title     = {War’s enduring effects on the development of egalitarian motivations and in-group biases},
  author    = {Bauer, Michal and Cassar, Alessandra and Chytilov{\'a}, Julie and Henrich, Joseph},
  journal   = {Psychological Science},
  volume    = {25},
  number    = {1},
  pages     = {47--57},
  year      = {2014},
  publisher = {Sage Publications Sage CA: Los Angeles, CA}
}

@article{bauer2016can,
  title     = {Can war foster cooperation?},
  author    = {Bauer, Michal and Blattman, Christopher and Chytilov{\'a}, Julie and Henrich, Joseph and Miguel, Edward and Mitts, Tamar},
  journal   = {Journal of Economic Perspectives},
  volume    = {30},
  number    = {3},
  pages     = {249--274},
  year      = {2016},
  publisher = {American Economic Association}
}

@article{bentzen2021crisis,
  title     = {In crisis, we pray: Religiosity and the {COVID}-19 pandemic},
  author    = {Bentzen, Jeanet Sinding},
  journal   = {Journal of Economic Behavior \& Organization},
  volume    = {192},
  pages     = {541--583},
  year      = {2021},
  publisher = {Elsevier}
}

@article{bentzen2019,
  title     = {Acts of God? Religiosity and Natural Disasters Across Subnational World Districts},
  author    = {Bentzen, Jeanet Sinding},
  journal   = {Economic Journal},
  volume    = {129},
  pages     = {2295-2321},
  year      = {2019},
}

@article{calo2017natural,
  title     = {Natural disasters and indicators of social cohesion},
  author    = {Calo-Blanco, Aitor and Kovářík, Jaromír and Mengel, Friederike and Romero, José Gabriel},
  journal   = {PLOS ONE},
  volume    = {12},
  number    = {6},
  pages     = {e0176885},
  year      = {2017},
  publisher = {Public Library of Science San Francisco, CA USA}
}

@article{campante2024virus,
  title   = {The Virus of Fear: The Political Impact of {Ebola} in the {United States}},
  author  = {Campante, Filipe and Depetris-Chauvin, Emilio and Durante, Ruben},
  journal = {American Economic Journal: Applied Economics},
  volume  = {16},
  number  = {1},
  pages   = {480--509},
  year    = {2024}
}

@article{cappelen2023universalism,
  title   = {Universalism: Global Evidence},
  author  = {Cappelen, Alexander W and Enke, Benjamin and Tungodden, Bertil},
  journal = {American Economic Review},
  volume  = {115},
  number  = {1},
  pages   = {43--76},
  year    = {2025},
 }

@article{deryuginamarx2021,
  title   = {Is the Supply of Charitable Donations Fixed? Evidence from Deadly Tornadoes},
  author  = {Deryugina, Tatyana and Marx, Benjamin M.},
  journal = {American Economic Review: Insights},
  volume  = {3},
  number  = {3},
  pages   = {383–398},
  year    = {2021},
 }

@article{enke2020moral,
  title     = {Moral values and voting},
  author    = {Enke, Benjamin},
  journal   = {Journal of Political Economy},
  volume    = {128},
  number    = {10},
  pages     = {3679--3729},
  year      = {2020},
  publisher = {The University of Chicago Press}
}

@article{enke2023structure,
  title     = {Moral Universalism and the Structure of Ideology},
  author    = {Enke, Benjamin and Rodr{\'\i}guez-Padilla, Ricardo and Zimmermann, Florian},
  journal   = {Review of Economic Studies},
  volume    = {90},
  number    = {4},
  pages     = {1934--1962},
  year      = {2023},
  publisher = {Oxford University Press}
}

@article{fridman2022increased,
  title     = {Increased generosity under {COVID}-19 threat},
  author    = {Fridman, Ariel and Gershon, Rachel and Gneezy, Ayelet},
  journal   = {Scientific Reports},
  volume    = {12},
  number    = {1},
  pages     = {4886},
  year      = {2022},
  publisher = {Nature Publishing Group}
}

@article{bhatia2025crisis,
  title={Crisis or Contentment? A Mixed-Method Exploration of Psychosocial Factors Influencing Meaning in Life During Midlife},
  author={Bhatia, Aditi and Laungani, Diksha and Hyland, Lynda and Shrivastava Kashi, Anita},
  journal={SAGE Open},
  volume={15},
  number={1},
  pages={21582440251323552},
  year={2025},
  publisher={SAGE Publications Sage CA: Los Angeles, CA}
}

@article{geiss2025inflation,
  title     = {Inflation of crisis coverage? {Tracking} and explaining the changes in crisis labeling and crisis news wave salience 1785-2020},
  author    = {Gei{\ss}, Stefan and Viehmann, Christina and Kelly, Conor A},
  journal   = {Journal of Communication},
  volume    = {75},
  number    = {1},
  pages     = {27--41},
  year      = {2025},
  publisher = {Oxford University Press}
}

@article{choi2007coevolution,
  title={The coevolution of parochial altruism and war},
  author={Choi, Jung-Kyoo and Bowles, Samuel},
  journal={science},
  volume={318},
  number={5850},
  pages={636--640},
  year={2007},
  publisher={American Association for the Advancement of Science}
}

@article{grimalda2021exposure,
  title     = {Exposure to {COVID}-19 is associated with increased altruism, particularly at the local level},
  author    = {Grimalda, Gianluca and Buchan, Nancy R and Pinate, Adriana and Urso, Giulia and Brewer, Marilynn B},
  journal   = {Scientific Reports},
  volume    = {11},
  number    = {1},
  pages     = {18950},
  year      = {2021},
  publisher = {Nature Publishing Group}
}

@article{meonverwimp2022,
  title     = {Pro-social behavior after a disaster: Evidence from a storm hitting an open-air festival},
  author    = {Meon, Pierre-Guillaume and Verwimp, Philip},
  journal   = {Journal of Economic Behavior and Organization},
  volume    = {198},
  pages     = {493-510},
  year      = {2022}
}

@article{schwirplies2023,
  title     = {Does additional demand for charitable aid increase giving? Evidence from Hurricane Sandy},
  author    = {Schwirplies, Claudia},
  journal   = {Journal of Economic Behavior and Organization},
  volume    = {209},
  pages     = {53-73},
  year      = {2023}
}

@article{enke2024universalism,
  title={Universalism and political representation: evidence from the field},
  author={Enke, Benjamin and Fisman, Raymond and Freitas, Luis Mota and Sun, Steven},
  journal={American Economic Review: Insights},
  volume={6},
  number={2},
  pages={214--229},
  year={2024},
  publisher={American Economic Association 2014 Broadway, Suite 305, Nashville, TN 37203}
}

@article{romano2021national,
  title     = {National parochialism is ubiquitous across 42 nations around the world},
  author    = {Romano, Angelo and Sutter, Matthias and Liu, James H and Yamagishi, Toshio and Balliet, Daniel},
  journal   = {Nature communications},
  volume    = {12},
  number    = {1},
  pages     = {4456},
  year      = {2021},
  publisher = {Nature Publishing Group UK London}
}

@article{scharf2022lift,
  title     = {Lift and shift: The effect of fundraising interventions in charity space and time},
  author    = {Scharf, Kimberley and Smith, Sarah and Wilhelm, Mark},
  journal   = {American Economic Journal: Economic Policy},
  volume    = {14},
  number    = {4},
  pages     = {407--443},
  year      = {2022},
  publisher = {American Economic Association}
}

@article{stewart2023group,
  title     = {Group reciprocity and the evolution of stereotyping},
  author    = {Stewart, Alexander J and Raihani, Nichola},
  journal   = {Proceedings of the Royal Society B},
  volume    = {290},
  number    = {1994},
  pages     = {20221834},
  year      = {2023},
  publisher = {The Royal Society}
}

@article{tessler2020anxiety,
  title     = {The anxiety of being {Asian American}: Hate crimes and negative biases during the {COVID}-19 pandemic},
  author    = {Tessler, Hannah and Choi, Meera and Kao, Grace},
  journal   = {American Journal of Criminal Justice},
  volume    = {45},
  pages     = {636--646},
  year      = {2020},
  publisher = {Springer}
}
%%%%%%%%%%%%%%%%%%%%%%%%%%%%%%%%%%%%%%%%%%%%%%%%%%%%%%%%%%%%%%%%%%%%%%%%%%%%%%%%%%%%
%%%%%%%%%%%%%%%%%%%%%%%%%%%%%%%%%%%%%%%%%%%%%%%%%%%%%%%%%%%%%%%%%%%%%%%%%%%%%%%%%%%%
%%%%%%%%%%%%%%%%%%%%%%%%%%%%%%%%%%%%%%%%%%%%%%%%%%%%%%%%%%%%%%%%%%%%%%%%%%%%%%%%%%%%
%%%%%%%%%%%%% FOLLOWING ARE ONLINE APPENDICES 
%%%%%%%%%%%%%%%%%%%%%%%%%%%%%%%%%%%%%%%%%%%%%%%%%%%%%%%%%%%%%%%%%%%%%%%%%%%%%%%%%%%%
%%%%%%%%%%%%%%%%%%%%%%%%%%%%%%%%%%%%%%%%%%%%%%%%%%%%%%%%%%%%%%%%%%%%%%%%%%%%%%%%%%%%
%%%%%%%%%%%%%%%%%%%%%%%%%%%%%%%%%%%%%%%%%%%%%%%%%%%%%%%%%%%%%%%%%%%%%%%%%%%%%%%%%%%%
\clearpage
\renewcommand{\appendixpagename}{Online Appendices}
\appendix
\appendixpage

\startcontents[sections]
\printcontents[sections]{l}{1}{\setcounter{tocdepth}{2}}

\numberwithin{equation}{section}
\counterwithin{figure}{section}
\counterwithin{table}{section}

%%%%%%%%%%%%%%%%%%%%%%%%%%%%%%%%%%%%%%%%%%%%%%%%%%%%%%%%%%%%%%%%%%%%%%%%%%%%%%%%%%%%
\clearpage
\section{Variables and Data Sources in this Study}
\label{asec:vardata}
%%%%%%%%%%%%%%%%%%%%%%%%%%%%%%%%%%%%%%%%%%%%%%%%%%%%%%%%%%%%%%%%%%%%%%%%%%%%%%%%%%%%

\begingroup 
\footnotesize
\begin{longtable}{L{0.19\textwidth} L{0.19\textwidth} L{0.42\textwidth} L{0.14\textwidth}}
\caption{Variables Matched to CAF Donations by Level} 
\label{tab:matched-caf} \\
\toprule
\textbf{Observation Level} & \textbf{Variable} & \textbf{Definition} & \textbf{Source} \\
\midrule
\endfirsthead

\toprule
\textbf{Observation Level} & \textbf{Variable} & \textbf{Definition} & \textbf{Source} \\
\midrule
\endhead

\midrule
\multicolumn{4}{r}{\footnotesize Continued on next page} \\
\bottomrule
\endfoot

\bottomrule
\endlastfoot

% ---------------- Donor level ----------------
Donor Level & Gender & Categories: Male, Female, or Uncategorized. & CAF \\
\midrule 

% ---------------- Postcode level ----------------
Postcode Level (15 households) & Wealth Index & Percentile rank of postcodes by property price. & HM Land Registry \\
\midrule 

% ---------------- Output Area ----------------
\multirow{7}{=}{Output Area (125 Households)} &
  Urban & Mark if the area has high residential-address density or overlaps an Amalgamated Built-Up Area with $\geq$\,10{,}000 residents. & Census 2021 \\
& Higher Education & Share of people with a Level 4 qualification or higher (degree-level or above). & Census 2021 \\
& Age 65 or Above & Share of residents aged 65 and over. & Census 2021 \\
& British White & Share of residents who are White British. & Census 2021 \\
& UK Identity & Share of people identifying with a UK identity. & Census 2021 \\
& Religious & Share of people identifying with a religion. & Census 2021 \\
\midrule 

% ---------------- Constituency ----------------
\multirow{13}{=}{Constituency (36,000 Households)} &
  Conservative Share & Average Conservative vote share across the 2015, 2017, and 2019 general elections. & UK Parliament \\
& Labour Share & Average Labour vote share across the 2015, 2017, and 2019 general elections. & UK Parliament \\
& Liberal Dem Share & Average Liberal Democrat vote share across the 2015, 2017, and 2019 general elections. & UK Parliament \\
& Local Belonging & Binary indicator for constituencies whose average sense of local belonging is at or above the median. & British Election Study \\
& Yesterday Anxiety & How anxious did you feel yesterday? (0-10). & British Election Study \\
& Yesterday Happiness & How happy did you feel yesterday? (0-10). & British Election Study \\
& Life Satisfaction & Overall, how satisfied are you with your life nowadays? (0-10). & British Election Study \\
& Life Worthwhile & Overall, to what extent do you feel that the things you do in your life are worthwhile? (0-10). & British Election Study \\
& Knows Covid Death & Has someone you personally know died of Covid-19? (rescaled to 0-1). & British Election Study \\
& Income Decreased & Has your household's income decreased since the Covid-19 outbreak? (rescaled to 0-1). & British Election Study \\
& Worried: Infection & Worry about you or someone close to you catching Covid-19 (rescaled to 0-1). & British Election Study \\
& Worried: Economy & Worry about the impact of Covid-19 on the economy (rescaled to 0-1). & British Election Study \\
& Worried: Life & Worry about the impact of Covid-19 on your way of life (rescaled to 0-1). & British Election Study \\
\midrule 

% ---------------- Upper Tier Local Authorities ----------------
\multirow{4}{=}{Upper Tier Local Authorities} &
  High Covid & Cumulative Covid-19 death rate up to 2022 (at or above median). & ONS \\
& Large GDP Shock & Post-Covid change in GDP per capita is at or below the median. & ONS \\
& Low Hate Crime & ULAD's average annual hate-crime rate (2015-2021) is at or above the median. & Belgioioso et al.\ (2023) \\
& Ethnic Homogeneity & Ethnic concentration, constructed with the Gini index over ULAD ethnic shares; higher values indicate less diversity. & Census 2021 \\

\end{longtable}
\endgroup

%%%%%%%%%%%%%%%%%%%%%%%%%%%%%%%%%%%%%%%%%%%%%%%%%%%%%%%%%%%%%%%%%%%%%%%%%%%%%%%%%%% 
\clearpage 
\section{Additional Context: Covid in England}
\label{asec:context}
%%%%%%%%%%%%%%%%%%%%%%%%%%%%%%%%%%%%%%%%%%%%%%%%%%%%%%%%%%%%%%%%%%%%%%%%%%%%%%%%%%%% 
\subsection{Covid Timeline in England}

\begin{figure}[!htbp]
    \centering
    \includegraphics[width=0.85\textwidth]{\figpath uk_covid_timeline_nonews.png}
    \caption{Covid timeline}
    \label{fig:timeline}
\end{figure}

\clearpage 
\subsection{Geographic Variation in Covid Exposure}

\begin{figure}[!htbp]
    \centering
    \includegraphics[width=0.75\textwidth]{\figpath ulad_high_low_covid.png}
    \caption{High versus Low Covid Exposure}
    \label{fig:ulad_high_low_covid}

    \vspace{5pt} 
    \parbox{\linewidth}{\footnotesize \textit{Notes}: The map highlights upper-tier local authorities with high versus low Covid exposure, where High Covid Exposure is defined as local authorities with cumulative Covid death rates up to the end of 2022 at or above the median across local authorities.}
\end{figure}

\begin{figure}[!htbp]
    \centering
    \includegraphics[width=0.75\textwidth]{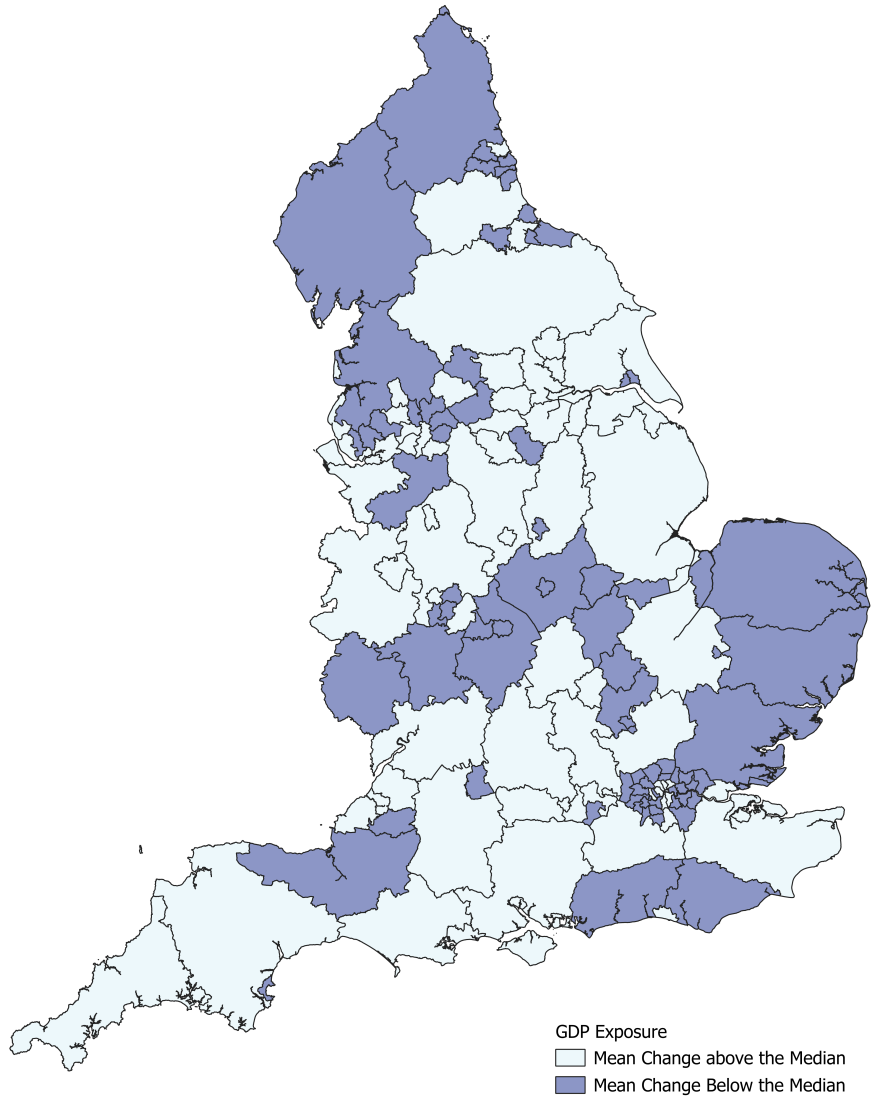}
    \caption{Large versus Small GDP Shock}
    \label{fig:ulad_high_low_gdp}

    \vspace{5pt} 
    \parbox{\linewidth}{\footnotesize \textit{Notes}: The map highlights upper-tier local authorities with large versus small GDP shocks, where Large GDP Shock is defined as local authorities with the change in mean annual GDP after Covid at or below the median.}
\end{figure}

\clearpage 
\subsection{Covid News in the UK}

\begin{figure}[!htbp]
    \centering
    \includegraphics[width=0.8\textwidth]{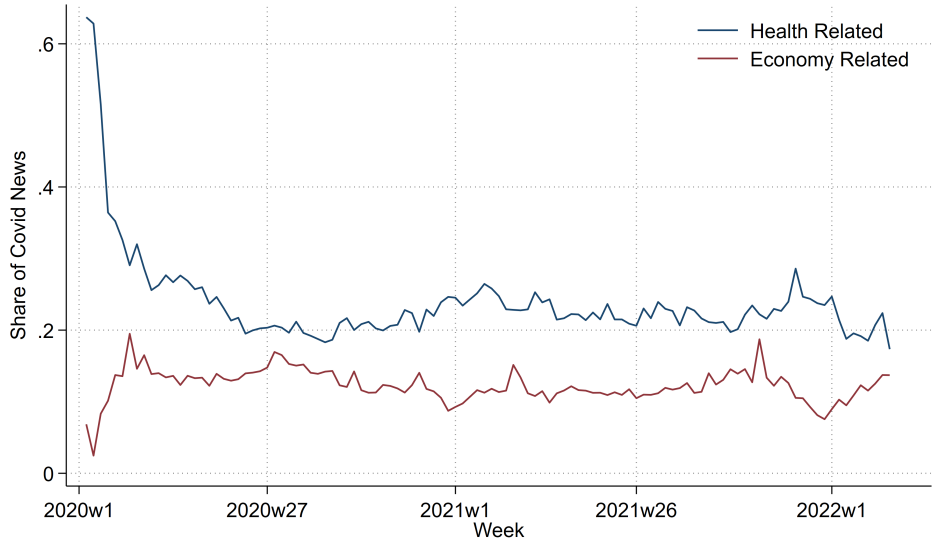}
    \caption{Share of Covid News by Topic}
    \label{fig:time_series_news_share_by_topic}

    \vspace{5pt} 
    \parbox{\linewidth}{\footnotesize \textit{Notes}: We use a sample of 524,514 news articles that mention Covid-19 (\enquote{covid}, \enquote{coronavirus}, or \enquote{corona}) from UK national newspapers collected via LexisNexis. We apply Latent Dirichlet Allocation topic modeling to extract 30 granular topics and then group them into three broad categories: health-related, economy-related, and others. The health category includes topics on testing and tracing, health care and communities, Covid cases and deaths, the health system, and vaccination. The economy category includes employment, business and companies, inflation, and the economy and market. The others category covers politics (domestic and foreign), sports, arts, and education-related restrictions, as well as general discussion.}
\end{figure}

\begin{figure}[!htbp]
    \centering
    \includegraphics[width=0.8\textwidth]{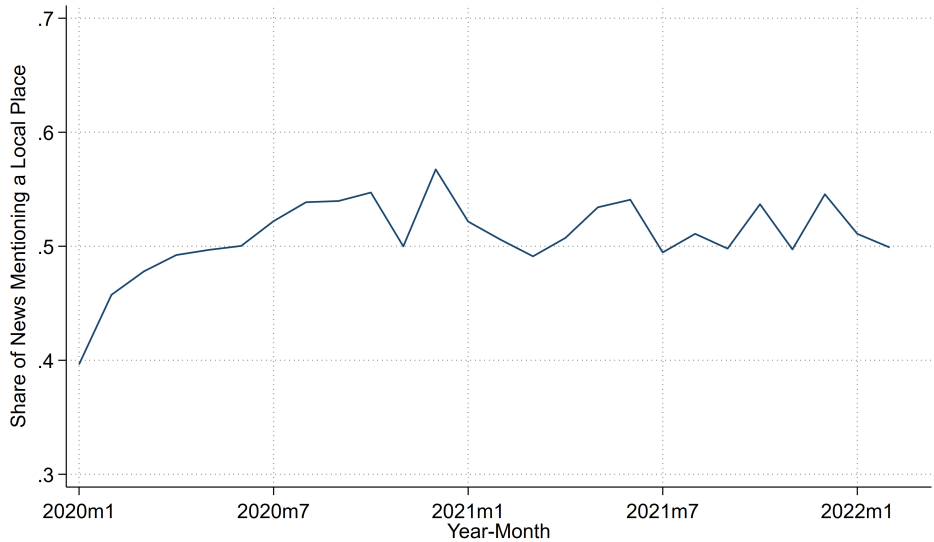}
    \caption{Share of Covid News Mentioning Local Places in the UK}
    \label{fig:time_series_local_share}

    \vspace{5pt} 
    \parbox{\linewidth}{\footnotesize \textit{Notes}: We geocode individual news articles by extracting locations and place names using named entity recognition (NER). We then match the extracted locations to country names or, where applicable, to Upper Tier Local Authorities (ULADs) in the UK using the Gazetteer of British Place Names. This figure plots the share of geocoded news articles that mention local places in the UK.}
\end{figure}

%%%%%%%%%%%%%%%%%%%%%%%%%%%%%%%%%%%%%%%%%%%%%%%%%%%%%%%%%%%%%%%%%%%%%%%%%%%%%%%%%%%%
\clearpage 
\section{Charitable Donations and Local Giving}
\label{asec:data}
%%%%%%%%%%%%%%%%%%%%%%%%%%%%%%%%%%%%%%%%%%%%%%%%%%%%%%%%%%%%%%%%%%%%%%%%%%%%%%%%%%%%

\subsection{Summary Statistics of Charity-level Data}

\begin{table}[!htbp]
    \centering
    \caption{Summary Statistics: Charity Commission Data}
    \label{tab:summary_stats_charity}
    \setlength{\tabcolsep}{18pt}          % Increases space between columns    
    \begin{tabular}{lrr}
        \toprule
        & \multicolumn{2}{c}{\textbf{Charity Group}} \\
        \cmidrule(lr){2-3}
        & \multicolumn{1}{c}{Local} & \multicolumn{1}{c}{Non-Local} \\
        \midrule

        \addlinespace[5pt]
        \textbf{Sample: All Charities} & & \\
        Mean yearly total income & 439,200 & 1,770,417 \\
        Median yearly total income & 48,256 & 77,142 \\
        Number of charities (average yearly) & 55,180 & 24,338 \\
        Share charities income $>$ £500k & 9.8\% & 19.2\% \\
        
        \addlinespace[10pt]
        \textbf{Sample: Charities with Income $>$ £500k} & & \\ 
        Mean yearly total income & 3,744,356 & 8,894,678 \\
        Median yearly total income & 1,263,615 & 1,574,260 \\
        Mean yearly total donations & 571,132 & 3,010,279 \\
        Median yearly total donations & 92,083 & 485,036 \\
        Number of charities (average yearly) & 5,421 & 4,680 \\
        \bottomrule
    \end{tabular}
    
    \vspace{5pt}
    \parbox{\linewidth}{\footnotesize \textit{Notes}: The data cover the years 2015–2022 and come from the annual reports submitted to the Charity Commission for England. Only charities with total annual income above £500,000 are required to report donation income; accordingly, this is the sample used in the regressions reported in Table \ref{tab:charity_level_local_vs_nonlocal}.}
\end{table}

\begin{table}[!htbp]
    \centering
    \caption{Charity Presence by Sectors: Local vs. Non-Local Charities}
    \label{tab:charity_presence_sectors}    
    \setlength{\tabcolsep}{20pt}         
    
    \begin{tabular}{lrr}
        \toprule
        \textbf{Sector} & \multicolumn{2}{c}{\textbf{Charity Group}} \\
        \cmidrule(lr){2-3}
        & \multicolumn{1}{c}{Local} & \multicolumn{1}{c}{Non-Local} \\
        \midrule
        Social Services             & 14.45 & 17.33 \\
        Religious                   & 11.40 & 13.99 \\
        Health/Medical Research     & 7.42  & 5.97  \\
        Education/Research/Culture  & 22.79 & 15.29 \\
        Environment                 & 3.38  & 3.61   \\
        Other                       & 40.56  & 43.81 \\
        \bottomrule
    \end{tabular}

    \vspace{5pt}
    \parbox{\linewidth}{\footnotesize \textit{Notes}: The table reports the shares of local and non-local charities by sector of main activity. Local and non-local charities are classified as described in Section \ref{sec:charity}: local charities operate exclusively at the local authority level, while non-local charities’ activities extend beyond this level, including regional, national, or international operations.}
\end{table}

\clearpage 
\subsection{Summary Statistics of Donor-level data from CAF}

\begin{figure}[!htbp]
    \centering
    \includegraphics[width=0.80\linewidth]{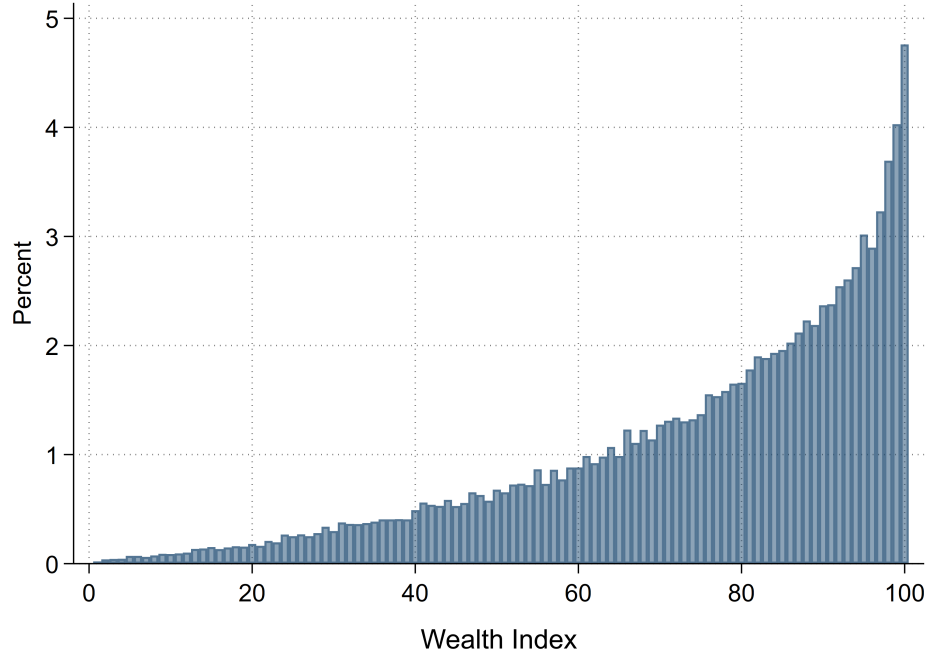}
    \caption{Wealth Distribution of CAF Donors' Postcodes}
    \label{fig:wealth_distribution}

    \vspace{5pt} 
    \parbox{\linewidth}{\footnotesize \textit{Notes:} The wealth index is defined as the percentile rank of the average property price in each postcode, calculated from transactions between 2000 and 2022 using data from the UK Land Registry. The figure shows the distribution of postcodes in which CAF donors reside across this wealth index. While CAF donors are represented across the entire distribution, postcodes in higher wealth percentiles are overrepresented: each percentile above the 60th accounts for more than 1\% of donor postcodes.}
\end{figure}

\begin{table}[!htbp]
    \centering
    \caption{Comparison of donation distribution between UKHLS and CAF}
    \label{tab:ukhls_caf}
    \begin{tabularx}{\textwidth}{
        >{\raggedleft\arraybackslash}X
        >{\raggedleft\arraybackslash}X 
        >{\raggedleft\arraybackslash}X}
        \toprule
        & \multicolumn{2}{c}{\textbf{Yearly Donation Size}} \\
        \cmidrule(lr){2-3}
        {} & UKHLS Donors & CAF Donors \\
        \midrule
        Mean   & 226    & 1,838   \\
        P1     & 4      & 20     \\
        P5     & 10     & 50    \\
        P10    & 12     & 90    \\
        P25    & 30     & 210    \\
        P50    & 96     & 544   \\
        P75    & 200    & 1,450   \\
        P90    & 500    & 3,500 \\
        P95    & 900    & 6,000 \\
        P99    & 2,510  & 19,804 \\        
        N      & 25,000 & 42,316 \\
        \bottomrule
    \end{tabularx}

    \vspace{5pt} 
    \parbox{\linewidth}{\footnotesize \textit{Notes}: This table compares the distribution of annual donation amounts between two groups: (1) individuals who reported positive charitable giving in the UK Household Longitudinal Study (UKHLS) in the wave that begins in 2014, and (2) donors in the Charities Aid Foundation (CAF) transaction data. In the UKHLS, respondents report total donations made to charities in the past 12 months. For comparability, annual donation amounts for CAF donors are constructed based on their donations in 2015. The table reports selected percentiles of the donation distribution and sample sizes for each group.}
\end{table}

\begin{table}[!htbp]
    \centering
    \caption{CAF Donors against the distribution of donors surveyed by UKHLS}
    \label{tab:donation_deciles}

    \begin{tabularx}{\linewidth}{
        >{\raggedright\arraybackslash}X 
        >{\raggedright\arraybackslash}X 
        >{\raggedright\arraybackslash}X 
        >{\raggedright\arraybackslash}X 
        >{\raggedright\arraybackslash}X}
        \toprule
        \multicolumn{4}{c}{Donors Surveyed by UKHLS} & 
        \multirow[c]{2}{=}{\raggedright Share of CAF Donors (\%)} \\
        \cmidrule(lr){1-4}
        Decile of Giving &
        Share of Donors (\%) &
        Minimum Donation (£) &
        Share of Total Giving (\%) & \\
        \midrule
        1  & 10 & 1   & 0.30 & 0.65 \\
        2  & 10 & 12  & 0.80 & 1.05 \\
        3  & 10 & 24  & 1.40 & 1.77 \\
        4  & 10 & 40  & 2.20 & 0.92 \\
        5  & 10 & 50  & 2.80 & 5.87 \\
        6  & 10 & 96  & 4.40 & 0.12 \\
        7  & 10 & 100 & 5.70 & 6.90 \\
        8  & 10 & 150 & 8.80 & 11.15 \\
        9  & 10 & 250 & 14.70 & 18.65 \\
        10 & 10 & 500 & 58.90 & 52.92 \\
        \bottomrule
    \end{tabularx}

    \vspace{5pt} 
    \parbox{\linewidth}{\footnotesize \textit{Notes}: This table divides UKHLS respondents who reported positive donations into ten equal-sized groups based on their annual donation amounts, using data from the 2014 wave. For each decile, the table reports the minimum annual donation amount in that group and the share of total donations accounted for by donors in that decile. The final column shows the distribution of CAF donors across these same decile thresholds, based on their annual donation amounts. This comparison illustrates how CAF donors are positioned within the broader population of donors as captured in the UKHLS survey.}
\end{table}

\begin{table}[!htbp]
    \centering
    \caption{Donation Shares by Donor Distance to Local Charity's Area of Operation}
    \label{tab:donation_shares_by_distance}
    \resizebox{\textwidth}{!}{\estauto{\tabpath distribution_of_distance_to_local_charities.tex}{15}{c}}

    \vspace{5pt} 
    \parbox{\linewidth}{\footnotesize \textit{Notes}: Local charities are defined as those operating exclusively at the upper-tier local authority level, as opposed to the regional, national, or international levels. This table reports the share of donations to local charities by the distance between their area of operation and their donors' location. Around 75\% of donations to local charities come from people living within 25 km of the charity's area of operation, and we define local donations as those made to local charities whose area of benefit is within 25 km of the donor's location.}
\end{table}

\begin{table}[!htbp]
    \centering
    \caption{Summary Statistics of CAF Donations}
    \label{tab:summary_statistics_caf_donation}
    \resizebox{\textwidth}{!}{\estauto{\tabpath summary_statistics_caf_donation.tex}{15}{c}}
    
    \vspace{5pt}
    \parbox{\linewidth}{\footnotesize \textit{Notes}: The data cover 2015--2022 and come from donor accounts managed by the Charities Aid Foundation. Our analysis focuses on CAF Charity Account holders and includes 44,436 charities that received donations during this period. This table presents statistics at the donor and donor-year levels.}
\end{table}

\clearpage
\subsection{Who Gives Locally?}

\begin{table}[!htbp]
    \centering
    \caption{Local Giving Before Covid}
    \label{tab:covariants_of_local_giving}
    \resizebox{0.75\textwidth}{!}{\estauto{\tabpath covariants_with_local_giving.tex}{15}{c}} 

    \vspace{5pt} 
    \parbox{\linewidth}{\footnotesize \textit{Notes}: This table presents the characteristics of individuals who donated to local charities before the Covid pandemic. Male is an indicator for donor gender. Wealth Index is the postcode-level property price percentile. Urban indicates urban areas. Higher Education is the share of residents with higher education or above. Age 65+ is the share of residents aged 65 and over. British White is the share of White British residents. Religious is the share of residents identifying with a religion. Self-claim UK Identity is the share of residents identifying with a UK identity. Conservative, Labour, and Liberal Democrat Vote Shares are averages across the 2015, 2017, and 2019 general elections. High Hate-Crime Area indicates upper-tier local authorities with average annual hate-crime rates (2015--2021) at or above the median. Ethnic Homogeneity is the Gini index of ethnic group shares (higher values indicate lower diversity). Sense of Local Belonging indicates the average level of local belonging at the constituency level. High Covid Exposure is defined as local authorities with a death rate at or above the median. Column (1) reports results from a linear probability model using all donors. Column (2) reports results from a linear probability model using donors who have made five or more donations. Column (3) reports results from a Tobit regression. Standard errors are clustered at the upper-tier local authority level.}
\end{table}

%%%%%%%%%%%%%%%%%%%%%%%%%%%%%%%%%%%%%%%%%%%%%%%%%%%%%%%%%%%%%%%%%%%%%%%%%%%%%%%%%%%%
\clearpage 
\section{Robustness Checks and Other Results}
\label{asec:check}
%%%%%%%%%%%%%%%%%%%%%%%%%%%%%%%%%%%%%%%%%%%%%%%%%%%%%%%%%%%%%%%%%%%%%%%%%%%%%%%%%%%%

\subsection{Charity Level Regression}

\begin{table}[!htbp]
    \centering
    \caption{Charity-Level Regressions: Local vs Non-Local with Imputed Donations}
    \label{tab:charity_level_imputed}
    \resizebox{0.95\textwidth}{!}{\estwide {\tabpath charity_commission_imputed.tex}{9}{c}} 

    \vspace{5pt} 
    \parbox{\linewidth}{\footnotesize \textit{Notes}: The dependent variable is the log of total donations received by charity $i$ (headquartered in ULAD $l$). When charity $i$ has total income below £500,000, and therefore does not report donation income, donations are imputed. `Local' is a dummy variable that equals 1 when charity $i$ operates exclusively at the local authority level. Robust standard errors clustered at the ULAD level in parentheses. $^{*}p<0.1$ $^{**}p<0.05$; $^{***}p<0.01$}
\end{table}

\clearpage 
\subsection{Changes on Local Giving Separately by Different Groups}

\begin{table}[!htbp]
    \centering
    \caption{Change in Local Giving after Covid by Covid Exposure}
    \label{tab:covid_on_local_giving}
    \resizebox{\textwidth}{!}{\estauto {\tabpath covid_on_local_giving.tex}{15}{c}} 

    \vspace{5pt} 
    \parbox{\linewidth}{\footnotesize \textit{Notes}: The outcome variables include a dummy variable indicating whether a donor gives to a local charity within 25 km of their location in a given year, the share of donations made to local charities by donation amount and by donation count. A year runs from March to February to align with the outbreak of the Covid pandemic. The data cover March 2015 to February 2022. \textit{Post Covid} indicates years since March 2020. \textit{High Covid} indicates upper-tier local authority districts (ULADs) with cumulative Covid death rates up to 2022 at or above the median; \textit{Low Covid} indicates those below the median. \textit{Year = 2020} and \textit{Year = 2021} separately indicate the first and second Covid years (March 2020--February 2021 and March 2021--February 2022). All regressions include donor fixed effects, year trends, sectoral composition controls, and total donations. Standard errors are clustered at the ULAD level. $^{}p<0.1$; $^{}p<0.05$; $^{}p<0.01$.}
\end{table}

\begin{table}[!htbp]
    \centering
    \caption{Change in Local Giving after Covid by Community Cohesion}
    \label{tab:community_on_local_giving}
    \resizebox{\textwidth}{!}{\estauto {\tabpath community_on_local_giving.tex}{15}{c}} 

    \vspace{5pt} 
    \parbox{\linewidth}{\footnotesize \textit{Notes}: The outcome variables include a dummy variable indicating whether a donor gives to a local charity within 25 km of their location in a given year, the share of donations made to local charities by donation amount and by donation count. A year runs from March to February to align with the outbreak of the Covid pandemic. The data cover March 2015 to February 2022. \textit{Post} indicates years since March 2020. High/Low Hate Crime and High/Low Ethnic Homogeneity are defined at the upper-tier local authority level using median splits across authorities: High Hate Crime indicates hate crime rates per 100,000 population at or above the median, and High Ethnic Homogeneity indicates ethnic homogeneity (measured using the Gini index of ethnic group shares) at or above the median; Low categories are defined as those below the median. High/Low Local Belonging is defined analogously at the constituency level, where High Local Belonging indicates constituencies with average local belonging at or above the median across constituencies, and Low Local Belonging indicates those below the median. All regressions include donor fixed effects, year trends, sectoral composition controls, and total donations. Standard errors are clustered at the ULAD level. $^{}p<0.1$; $^{}p<0.05$; $^{}p<0.01$.}
\end{table}

\begin{table}[!htbp]
    \centering
    \caption{Change in Local Giving after Covid by GDP Shock}
    \label{tab:gdp_on_local_giving}
    \resizebox{\textwidth}{!}{\estauto {\tabpath gdp_on_local_giving.tex}{15}{c}} 

    \vspace{5pt} 
    \parbox{\linewidth}{\footnotesize \textit{Notes}: The outcome variables include a dummy variable indicating whether a donor gives to a local charity within 25 km of their location in a given year, the share of donations made to local charities by donation amount and by donation count. A year runs from March to February to align with the outbreak of the Covid pandemic. The data cover March 2015 to February 2022. \textit{Post} indicates years since March 2020. \textit{Large GDP Shock} indicates ULADs with the change in mean annual GDP after Covid at or below the median; \textit{Small GDP Shock} indicates those above. All regressions include donor fixed effects, year trends, sectoral composition controls, and total donations. Standard errors are clustered at the ULAD level. $^{}p<0.1$; $^{}p<0.05$; $^{}p<0.01$.}
\end{table}

\clearpage
\subsection{Additional Outcome, Treatment and Specifications}

\begin{figure}[!htbp]
    \centering
    \includegraphics[width=0.75\linewidth]{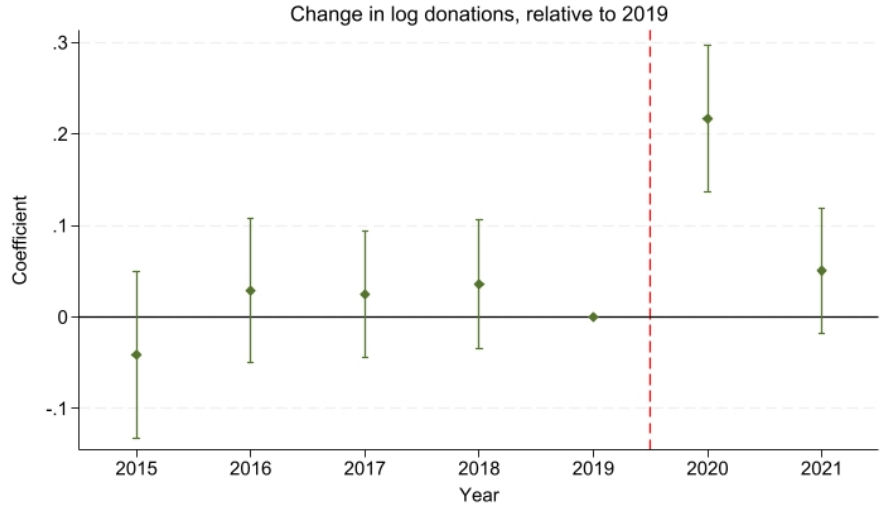}
    \caption{Giving to Local versus Non-local Charities: Charity Level Evidence}
    \label{fig:event_cc}

    \vspace{5pt} 
    \parbox{\linewidth}{\footnotesize \textit{Notes:} This figure illustrates the differential change in donations to local versus non-local charities, defining local charities as those that operate exclusively at the local authority level. The event study uses 2019 as the base year. The underlying regressions incorporate charity fixed effects, year fixed effects, and controls for both fundraising expenditure and the GDP per capita of the local authority where the charity is headquartered. Standard errors are clustered at the ULAD level. Point estimates are plotted alongside their 95\% confidence intervals.}
\end{figure}

\begin{figure}[!htbp]
    \centering
    \includegraphics[width=0.75\linewidth]{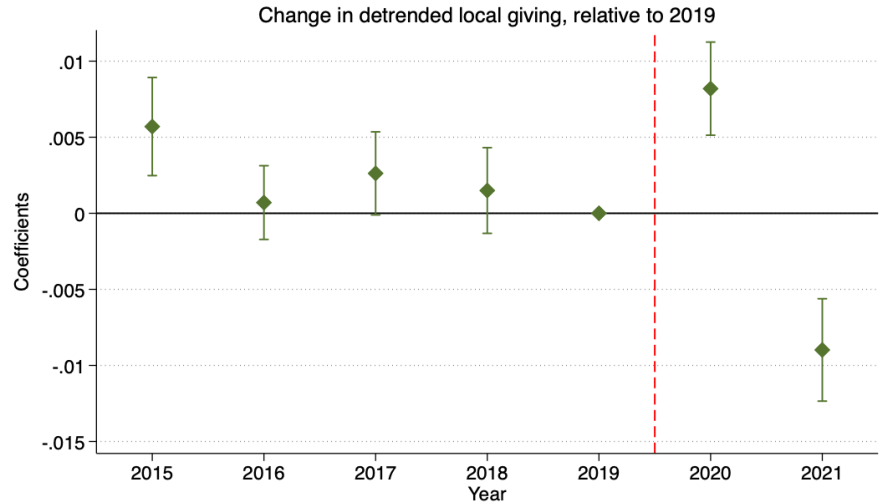}
    \caption{Changes in Local Giving Following Covid-19: Donor Level Evidence}
    \label{fig:event_study_year_only}

    \vspace{5pt} 
    \parbox{\linewidth}{\footnotesize \textit{Notes:} We first remove the linear trend from the change in local giving and then run a regression of the de-trended local giving on year dummies, using 2019 (the year before Covid-19) as the baseline. We control for donor fixed effects, the sectoral composition of donations, and total donation value, as in the main Table \ref{tab:differential_response_to_decappeal}. Standard errors are clustered at the ULAD level. The figure plots the point estimates with their 95\% confidence intervals.}
\end{figure}

\begin{figure}[!htbp]
    \centering
    \includegraphics[width=0.75\linewidth]{\figpath event_study_high_covid.png}
    \caption{Changes in Local Giving Following COVID-19: High versus Low Exposure}
    \label{fig:event_study_high_covid}

    \vspace{5pt} 
    \parbox{\linewidth}{\footnotesize \textit{Notes:} I run an event study using 2019 as the baseline year, comparing the change in local giving in high-COVID areas relative to areas with lower COVID exposure. I control for donor fixed effects, the sectoral composition of donations, and total donation value. High COVID indicates upper-tier local authority districts (ULADs) with cumulative COVID death rates up to 2022 at or above the median. In additional specifications, I interact the post-COVID indicator with gender and several area characteristics, including large GDP shocks, strong local belonging, wealth, urbanisation, the share of people with higher education, the share claiming UK identity, the share with a religious affiliation, the share aged 65 or above, the share identifying as British White, and Conservative vote share. Standard errors are clustered at the ULAD level. The figure plots the point estimates together with their 95\% confidence intervals.}
\end{figure}

\clearpage 
\begin{table}[!htbp]
    \centering
    \caption{Donation Response to Continuous Deaths Rate}
    \label{tab:main_diff_change_after_covid_contT}
    \resizebox{\textwidth}{!}{\estauto{\tabpath main_diff_change_after_covid_contT.tex}{15}{c}} 

    \vspace{5pt} 
    \parbox{\linewidth}{\footnotesize \textit{Notes:} The dependent variable is share of donations an individual makes to local charities within 25km. A year runs from March to February to align with the outbreak of the Covid pandemic. The data cover March 2015 to February 2022. \textit{Post Covid} indicates years since March 2020. Instead of using a dummy variable for high/low Covid exposure, we use the raw, continuous measure of the cumulative Covid death rate by the end of 2022. \textit{Large GDP Shock} indicates ULADs with the change in mean annual GDP after Covid at or below the median. \textit{Strong Local Belonging} indicates constituencies where the mean sense of belonging to the local community is at or above the median. All regressions include donor fixed effects, year trends, sectoral composition controls, total donations, and interactions of \textit{Post Covid} with gender, wealth, urbanization, the shares of people with higher education, claiming UK identity, with a religious affiliation, aged 65 or above, identifying as British White, and Conservative vote share. Column (4) includes region by year fixed effects; Column (6) runs the regression using only observations with a positive donation amount. \textit{Year = 2020} and \textit{Year = 2021} in Column (5) separately capture the first and second Covid years. Standard errors are clustered at the ULAD level. $^{*}p<0.1$ $^{**}p<0.05$; $^{***}p<0.01$.}
\end{table}

\begin{table}[!htbp]
    \centering
    \caption{Change in Share of Giving to Different Causes after Covid}
    \label{tab:change_in_sectoral_giving}
    \resizebox{\textwidth}{!}{\estauto {\tabpath befaft_on_sectoral_giving.tex}{15}{c}} 

    \vspace{5pt} 
    \parbox{\linewidth}{\footnotesize \textit{Notes}: The outcome variable is the share of donations a donor makes to charities in different sectors, namely social services, religious, health, education-research-and-culture, and other charities. A year runs from March to February to align with the outbreak of the Covid pandemic. The data cover March 2015 to February 2022. \textit{Post Covid} indicates years since March 2020. All regressions include donor fixed effects, year trends, and total donations. Standard errors are clustered at the ULAD level. $^{}p<0.1$; $^{}p<0.05$; $^{}p<0.01$.}
\end{table}

\begin{table}[!htbp]
    \centering
    \caption{Change in Local Giving after Covid by Charity Cause}
    \label{tab:change_in_local_giving_by_cause}
    \resizebox{\textwidth}{!}{\estauto {\tabpath befaft_on_local_giving_by_cause.tex}{15}{c}} 

    \vspace{5pt} 
    \parbox{\linewidth}{\footnotesize \textit{Notes}: The outcome variable is the share of donations a donor makes to local charities with 25 km of their location. A year runs from March to February to align with the outbreak of the Covid pandemic. The data cover March 2015 to February 2022. \textit{Post Covid} indicates years since March 2020. The table presents results conditional on donations to a specific sector, namely social services, religious, health, education, research and culture, and other charities. The regressions include donor fixed effects, year trends, and total donations. Standard errors are clustered at the ULAD level. $^{}p<0.1$; $^{}p<0.05$; $^{}p<0.01$.}
\end{table}

\begin{table}[!htbp]
    \centering
    \caption{Change in local giving after Covid using an alternative distance threshold}
    \label{tab:robustness_distance_thresholds}
    \resizebox{\textwidth}{!}{\estauto{\tabpath robustness_distance_thresholds.tex}{15}{c}} 

    \vspace{5pt} 
    \parbox{\linewidth}{\footnotesize \textit{Notes:} The dependent variable is the share of local donations. Local donations are defined as donations made to local charities within a given distance(i.e., 10 km, 25 km, and 50 km). A year runs from March to February to align with the outbreak of the Covid pandemic. The data cover March 2015 to February 2022. \textit{Post Covid} indicates years since March 2020. \textit{High Covid} indicates upper-tier local authority districts (ULADs) with cumulative Covid death rates up to 2022 at or above the median. \textit{Large GDP Shock} indicates ULADs with the change in mean annual GDP after Covid at or below the median. \textit{Strong Local Belonging} indicates constituencies where the mean sense of belonging to the local community is at or above the median, and interactions of \textit{Post Covid} with gender, wealth, urbanization, the shares of people with higher education, claiming UK identity, with a religious affiliation, aged 65 or above, identifying as British White, and Conservative vote share. Standard errors are clustered at the ULAD level. Significance levels are denoted by $p < 0.10$ (*), $p < 0.05$ (**), and $p < 0.01$ (***).}
\end{table}

% \begin{table}[!htbp]
%     \centering
%     \caption{Monthly Donation Response to Real-time Deaths Rate}
%     \label{tab:befaft_interaction_results}
%     \resizebox{\textwidth}{!}{\estauto {\tabpath befaft_interaction_results.tex}{15}{c}} 

%     \vspace{5pt} 
%     \parbox{\linewidth}{\footnotesize \textit{Notes:} The dependent variable is the share of donations allocated to local charities within 25km, aggregated at the ULAD-month level. Columns (1) and (2) measure the share by value, and columns (3) and (4) measure the share by number. \textit{Post Covid} indicates months starting from March 2020. \textit{Death Rate} denotes the Covid-19 death rate in percentage at the ULAD level, specific to the month. \textit{March-April-May} is an indicator for those specific months. All regressions include ULAD fixed effects, month fixed effects, time trends, total donations, and differential effects by socio-demographic features (gender, age, education, wealth, British White, UK identity, and Conservative vote share). Standard errors are clustered at the ULAD level. $^{*}p<0.1$ $^{**}p<0.05$; $^{***}p<0.01$.}
% \end{table}

\clearpage 
\subsection{Change in Local Giving Following Lockdowns}
\label{sec:lockdowns}

The following figure suggests that there is a change in local giving in response to local need. But, people become less sensitive to local need after the initial shock.

\begin{figure}[!htbp]
    \centering
    \includegraphics[width=0.9\textwidth]{\figpath national_lockdown_on_local_nat.png}
    \caption{Change in Local Giving Following National Lockdowns}
    \label{fig:national_lockdown_on_local_nat}

    \vspace{5pt} 
    \parbox{\linewidth}{\footnotesize \textit{Notes}: The data are aggregated at the weekly level across places in high-Covid and low-Covid areas. High-Covid areas are upper-tier local authority districts (ULADs) with cumulative Covid death rates up to 2022 at or above the median. The figure reports event-study coefficients from the regression of the share of local giving on leads and lags around the announcement of the first national lockdown (week 12 of 2020; 23 March 2020) and the second national lockdown (week 44 of 2020; 31 October 2020), separately for high- and low-Covid-exposure areas. The specification includes week fixed effects and high-Covid-area fixed effects, and controls for sectoral giving shares. Standard errors are clustered at the week-by-(high/low)-area level.}
\end{figure}

\begin{figure}[!htbp]
    \centering
    \includegraphics[width=0.9\textwidth]{\figpath national_lockdown_on_local_nat_deathcontrol.png}
    \caption{Change in local giving following national lockdowns (controls for deaths)}
    \label{fig:national_lockdown_on_local_nat_deathcontrol}

    \vspace{5pt}
    \parbox{\linewidth}{\footnotesize \textit{Notes}: The data are aggregated at the weekly level across places in high-Covid and low-Covid areas. High-Covid areas are upper-tier local authority districts (ULADs) with cumulative Covid death rates up to 2022 at or above the median. The figure reports event-study coefficients from the regression of the share of local giving on leads and lags around the announcement of the first national lockdown (week 12 of 2020; 23 March 2020) and the second national lockdown (week 44 of 2020; 31 October 2020), estimated separately for high- and low-Covid-exposure areas. The specification includes week fixed effects and high-Covid-area fixed effects, and controls for sectoral giving shares and weekly deaths in high- and low-Covid areas. Standard errors are clustered at the week-by-(high/low)-area level.}
\end{figure}

\begin{figure}[!htbp]
    \centering
    \includegraphics[width=0.95\textwidth]{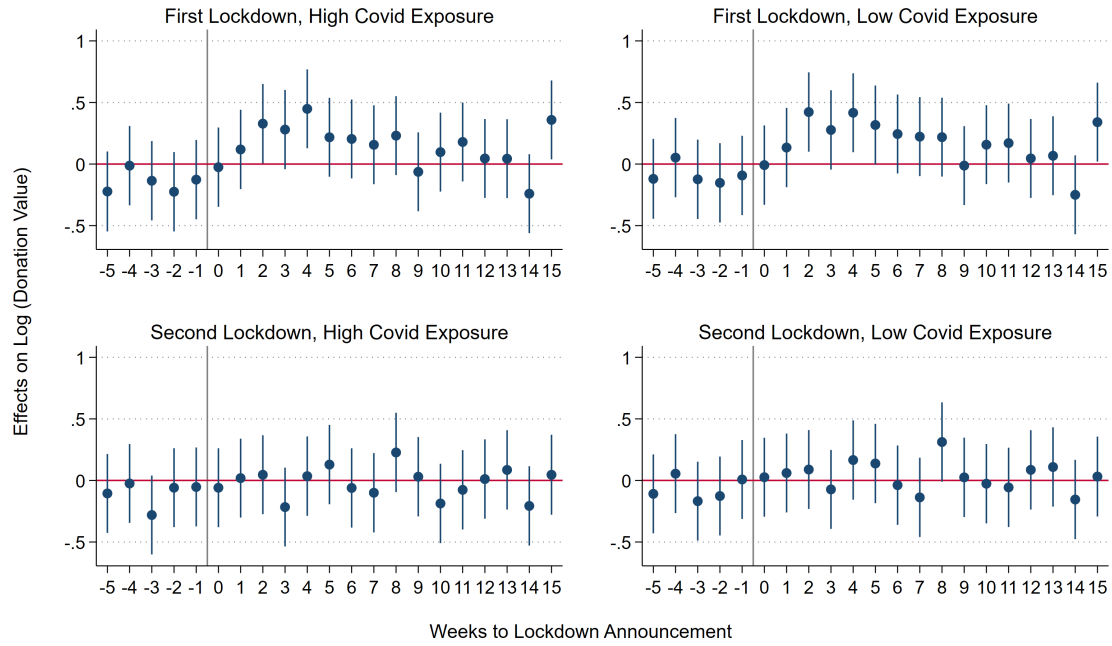}
    \caption{Change in donation value following national lockdowns}
    \label{fig:national_lockdown_on_donval}

    \vspace{5pt}
    \parbox{\linewidth}{\footnotesize \textit{Notes}: The data are aggregated at the weekly level across places in high-Covid and low-Covid areas. High-Covid areas are upper-tier local authority districts (ULADs) with cumulative Covid death rates up to 2022 at or above the median. The figure reports event-study coefficients from the regression of weekly total donation on leads and lags around the announcement of the first national lockdown (week 12 of 2020; 23 March 2020) and the second national lockdown (week 44 of 2020; 31 October 2020), separately for high- and low-Covid-exposure areas. The specification includes week fixed effects and high-Covid-area fixed effects, and controls for sectoral giving shares. Standard errors are clustered at the week-by-(high/low)-area level.}
\end{figure}

\begin{figure}[!htbp]
    \centering
    \includegraphics[width=0.98\textwidth]{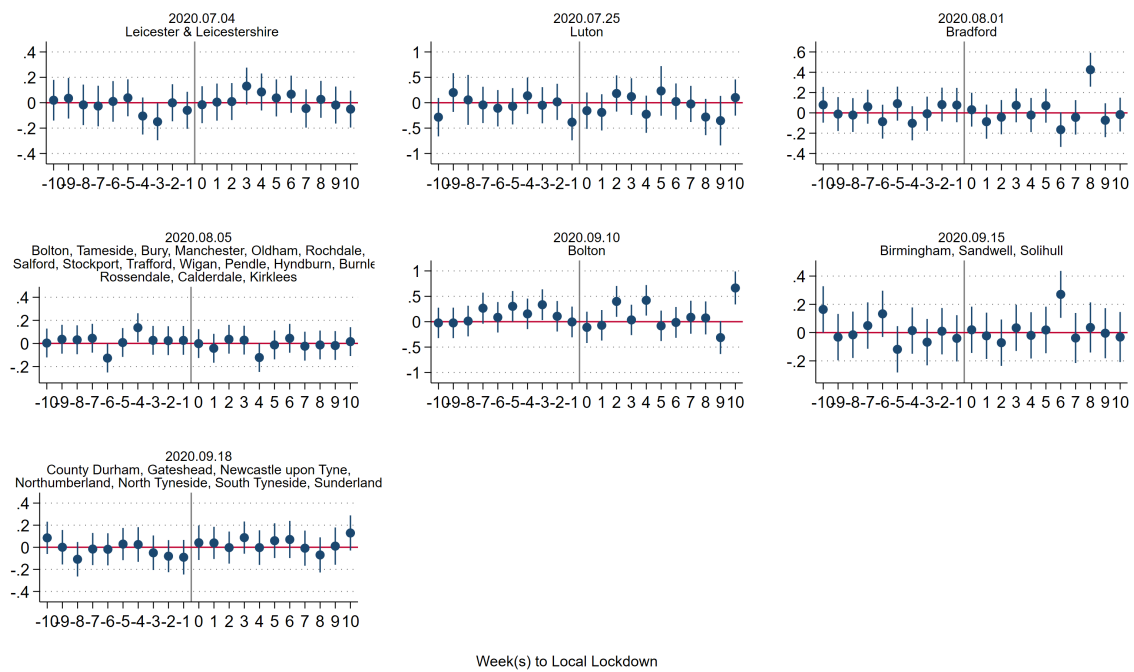}
    \caption{Change in Local Giving Following Local Lockdowns}
    \label{fig:change_in_local_giving_following_local_lockdowns}

    \vspace{5pt}
    \parbox{\linewidth}{\footnotesize \textit{Notes}: Donations are aggregated weekly at the upper-tier local authority level. For each local lockdown, I run the regression using only donations from the affected local area. The figure reports event-study coefficients from the regression of the share of local giving on leads and lags around the announcement of the local lockdown. The specification includes week fixed effects and controls for the first and second national lockdowns. Standard errors are clustered at the week-by-(high/low)-area level.}
\end{figure}

\end{document}